\documentclass[preprint,preprintnumbers,amsmath,amssymb, nofootinbib]{revtex4}

\usepackage{graphicx}
\usepackage{dcolumn}
\usepackage{bm}
\usepackage{epsfig}

\usepackage{amsmath}
\usepackage{amssymb}

\def\fun#1#2{\lower3.6pt\vbox{\baselineskip0pt\lineskip.9pt
  \ialign{$\mathsurround=0pt#1\hfil##\hfil$\crcr#2\crcr\sim\crcr}}}

\input epsf

\newcommand{\be}{\begin{equation}}
\newcommand{\ee}{\end{equation}}
\newcommand{\bea}{\begin{eqnarray}}
\newcommand{\eea}{\end{eqnarray}}

\newcommand{\lsim}{\mathrel{\raisebox{-.6ex}{$\stackrel{\textstyle<}{\sim}$}}}

\begin{document}
\global\long\def\met{\not{\!{\rm E}}_{T}}

\vspace*{1cm}

\title{Singlet scalars as Higgs imposters at the Large Hadron Collider}

\vspace*{0.2cm}

\author{
\vspace{0.5cm} 
Ian Low$^{a,b}$, Joseph Lykken$^{c}$, and Gabe Shaughnessy$^{a,b}$ }
\affiliation{
\vspace*{.2cm}
$^a$ \mbox{High Energy Physics Division, Argonne National Laboratory, Argonne, IL 60439}\\
$^b$ \mbox{Department of Physics and Astronomy, Northwestern University, Evanston, IL 60208} \\
$^c$  \mbox{Fermi National Accelerator Laboratory, P.O. Box 500, Batavia, IL 60510}
\vspace*{0.8cm}}

\begin{abstract}
\vspace*{0.5cm}
An electroweak singlet scalar can couple to pairs of vector bosons
through loop-induced dimension five operators.
 Compared to a Standard Model Higgs boson, the singlet decay widths in the  diphotons and $Z\gamma$ channels
are generically enhanced, while decays into massive final states like $WW$ and $ZZ$ are kinematically disfavored. 
The overall event rates into $\gamma\gamma$ and $Z\gamma$ can exceed the Standard Model expectations by orders of magnitude.  
Such a singlet may appear as a resonant signal in the $\gamma\gamma$ and $Z\gamma$ channels, even with a mass
above the $WW$ kinematic threshold.

\end{abstract}


\maketitle

\section{Introduction}\label{sec:into}
Experiments at the CERN Large Hadron Collider are collecting data from proton-proton collisions at
$\sqrt{s} = 7$ TeV and examining the experimental signatures for the production and decay
of the Higgs boson particle predicted by the Standard Model (SM). Both the production cross section
and decay branching fractions of the SM Higgs $h$ can be accurately computed as a function of the
unknown Higgs mass. These quantities, combined with considerations of SM backgrounds and
detector resolutions for the relevant final states, motivates a search strategy that focuses on the
diphoton decay $h\to \gamma\gamma$ for Higgs mass $\lsim$ 130 GeV$/c^2$, and for heavier Higgs
the decay into two massive vector bosons $h\to W^+W^-$ or $h\to ZZ$.

A resonant signal in these diboson channels does not however constitute the discovery of
the SM Higgs boson. The true dynamical mechanisms of electroweak symmetry breaking
and fermion mass generation are unknown, and may involve a variety of new heavy particles carrying SM
charges and/or exotic quantum numbers. These may include heavy bosons of spin 0, 1 or larger that
can be resonantly produced at the LHC. To the extent that these bosons are part of an ``extended
Higgs sector" responsible for electroweak symmetry breaking and/or fermion mass generation,
it is natural to assume that one or more may be relatively light.

Thus the first resonant diboson signal observed at the LHC may not originate from a SM
 Higgs. If the observed event rate for this resonance is compatible, within experimental
uncertainties, to the rate predicted for the Higgs, we must still confront the problem of whether 
we have observed the SM Higgs or a look-alike \cite{Cao:2009ah,Gao:2010qx, DeRujula:2010ys}.
If the event rate is much larger than that
predicted for the Higgs, we then need to address the problem of determining the true identity 
and significance of
this Higgs imposter.\footnote{Here we have used {\it imposter} to denote a bosonic resonance
that can cause an excess in a canonical Higgs search but is clearly distinguished from a SM Higgs by
a significantly nonstandard rate in one or more channels; while {\it look-alike} denotes a more
difficult case requiring knowledge of the correlations in the visible decay products of the final state.}
Within the normal confines of quantum field theory one can classify all possible 
Higgs imposters and look-alikes decaying in one or more diboson channels. Such a
classification uses the spin and $CP$ properties of the new particle, its transformation
properties under both the electroweak gauge group and the SM custodial symmetry,
its dominant couplings to various pairs of vector bosons (including two gluons), and its couplings to fermions.

As already noted in Ref.~\cite{Low:2010jp}, a particularly interesting candidate for a Higgs imposter
is a massive spin zero particle transforming as a SM singlet. (See also Refs.~\cite{Fox:2011qc, Sato:2011gp}.)
 Such singlets are 
common in theories with extended Higgs sectors, and their couplings to quarks and leptons are naturally
suppressed by some combination of SM Yukawa couplings, mixing angles, and ratios of
scalar vacuum expectation values (VEV) \cite{Lykken:2008bw}.
While distinguishing generic scalar models using both gauge boson and 
fermion couplings can be worthwhile~\cite{Barger:2009me}, in this paper we will assume that the couplings to SM
fermions are negligible, and focus instead on the dominant couplings to dibosons.  

In section \ref{sec:diphoton} we review the various possible mechanisms to
generate a diphoton resonance, which is the primary discovery channel in the low Higgs mass region, 
 with an event rate enhanced compared to that of a
 Standard Model Higgs. In section \ref{sec:demo} we consider singlet scalars
coupling through dimension five operators to dibosons, and observe a natural
hierarchy of decay widths favoring $gg$, $\gamma\gamma$ and $Z\gamma$ 
over $W^+W^-$ and $ZZ$. In section \ref{sec:pheno} we estimate the integrated
luminosity required in 7 TeV or 14 TeV running at the LHC to discover or 
exclude these Higgs imposters, followed by the conclusion in section V.

\section{Enhancing diphoton resonances}\label{sec:diphoton}

For models with a neutral scalar sitting in an $SU(2)_L$ doublet like the Higgs boson, it is difficult to enhance the event rate 
into the diphoton decay channel by an order of magnitude. The reason is 1)  the photon is massless, which implies that the
Higgs can only couple to the photon through dimension five operators, and 2) the Higgs does not carry electric charge
so the coupling is loop-induced. The leading order operator is
\be
\label{eq:cgamma}
a_\gamma \frac{\alpha_{em}}{4\pi} \frac{h}{v} F_{\mu\nu} F^{\mu\nu} \ ,
\ee
where $v=(\sqrt{2} G_F)^{-1/2}\approx 246$ GeV and $a_\gamma$ is, in general, an order unity constant. 
On the other hand, since the VEV of the Higgs boson gives masses to $W$ and $Z$ bosons, couplings to 
$WW$ and $ZZ$  are contained in the Higgs kinetic term:
\be 
\label{eq:wzmassterm}
\frac12 g^2 v\, h W_\mu^-W^{+\, \mu} + \frac14 \frac{g^2}{c_w^2} v\, h Z_\mu Z^\mu \ .
\ee
Therefore for $m_h\agt 120$ GeV$/c^2$  the decay partial width into $WW$ is at least two orders of magnitude larger 
than that  into $\gamma\gamma$, while at $m_h=115$ GeV$/c^2$  the $WW$ partial width is larger 
by a factor of 30 \cite{Djouadi:2005gi}, even though for these masses the $WW$ channel is kinematically suppressed.
However, we do not measure partial widths directly but only the event rate which is the production cross section times branching ratio:
\be
B\sigma (gg\to h\to V_1V_2) = \sigma (gg\to h)\times Br(h\to V_1V_2) \ ,
\ee
from which we see that $B\sigma$ can be enhanced by increasing either the production cross section, decay branching ratio, or both. 

One possibility of achieving a larger branching fraction in diphoton mode is to decrease the total decay width of the Higgs by turning off its couplings to fermions, the $b$-quark in particular, which results in an increase in the branching ratio by
\be
\frac{1}{1-Br(h\to f\bar{f})} \ .
\ee
Using the branching fractions for a SM Higgs boson, the largest enhancement that 
can be achieved this way is for $m_h=115$ GeV$/c^2$, 
by a factor of 6.3 when couplings to {\em all} fermions are turned off. The increase is only a factor of 3.8 if just the 
coupling to the $b$-quark is turned off. For higher masses the $WW$ mode becomes increasingly dominant and the 
enhancement due to turning off fermionic couplings correspondingly even smaller. 

\begin{figure}
\includegraphics[scale=0.55, angle=0]{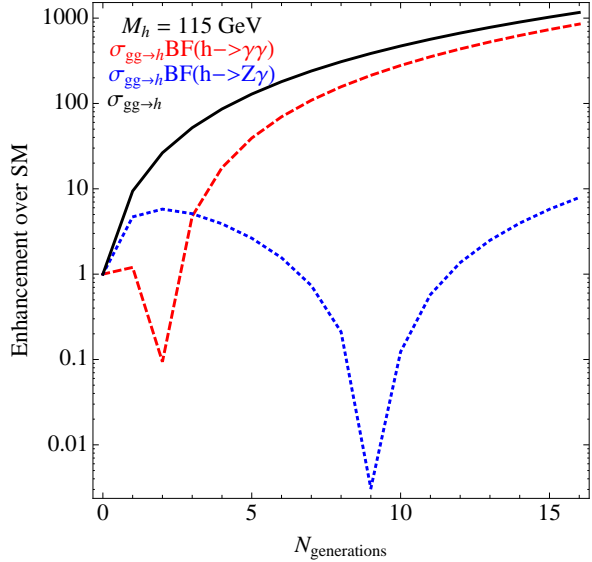}  
\caption{\label{fig:1}{\em Enhancement in the event rate of a 115 GeV$/c^2$ Higgs boson versus the number of extra generations. Each generation includes one quark doublet and one lepton doublet.}
}
\end{figure}

Other possibilities include enhancing couplings to two gluons and two photons, which increase the production rate of the Higgs and partial width into diphotons, respectively. Since both types of couplings are loop-induced, these possibilities can be realized by introducing new heavy particles running in the loop. However, effects of a new particle generically decouple as its mass becomes heavy and scale like $v^2/m_{\rm new}^2$ with respect to SM expectations. So a large enhancement generally requires very light masses below the scale of $v\sim 246$ GeV/$c^2$. The only exception is when there is a fourth generation fermion, which does not decouple \cite{D'Hoker:1984pc}.
A fourth generation interferes constructively with the top 
quark and increases the Higgs production rate by roughly a factor of $3^2=9$. Nonetheless, the same fourth 
generation interferes destructively with the $W$ boson loop contributing to Eq.~(\ref{eq:cgamma}), resulting in
a net decrease of 
the partial decay width into $\gamma\gamma$. So a large multiplicity of extra generations is required 
to enhance $B\sigma$ in the $\gamma\gamma$ channel by an order of magnitude, as can be seen in Fig.~\ref{fig:1}.\footnote{When adding a fourth generation to two-Higgs-doublet models such as supersymmetry, it is possible to enhance significantly the branching fraction of the pseudo-scalar Higgs  $A\to \gamma\gamma$ \cite{Cotta:2011bu,Gunion:2011ww}.}  Recent studies on the impact of fourth generations on the SM Higgs phenomenology can be found in Refs.~\cite{Kribs:2007nz, Ruan:2011qg}.

In the next section we demonstrate that an electroweak singlet scalar 
suffers from none of the issues above, and can naturally have a large event rate in the diphoton
and $Z\gamma$ channels.

\section{The singlet is democratic}\label{sec:demo}

To be specific, let's write down the effective operators that couple a singlet scalar $S$ to pairs of vector bosons. There are only three of them at leading order:
\be
\label{eq:Sope}
{\cal L}_{eff}= \kappa_{g} \frac{\alpha_s}{4\pi} \,\frac{S}{4 m_S} G_{\mu\nu}^a G^{a\, \mu\nu} +
\kappa_{W} \frac{\alpha_{em}}{4\pi s_w^2} \,\frac{S}{4 m_S} W_{\mu\nu}^a W^{a\, \mu\nu} +
\kappa_{B} \frac{\alpha_{em}}{4\pi c_w^2} \,\frac{S}{4 m_S} B_{\mu\nu} B^{\mu\nu} \ ,
\ee 
where $s_w$ and $c_w$ are the sine and cosine of the Weinberg angle. We have also chosen to normalize the
effective operators to the scalar mass. Naive dimensional analysis 
suggests $\kappa_g \sim \kappa_W \sim \kappa_B \sim {\cal O}(m_S/\Lambda)$, 
where  $\Lambda$ is the mass scale of new physics. The three operators control the 
partial decay widths of $S$ into all five possible pairs of vector bosons: 
$V_1V_2=\{WW, ZZ, Z\gamma, \gamma\gamma, gg\}$.  On general grounds, when $m_S$ is below the $2m_W$ threshold, 
we expect that decays into massless states such as $gg$ and $\gamma\gamma$ 
are kinematically preferred over massive final states. At higher masses, the $WW$ mode 
should become a viable decay channel since the $SU(2)_L$ coupling is slightly stronger
 than the $U(1)_{em}$ one, and because $WW$ encompasses two gauge eigenstates $W^1$ and $W^2$.

In terms of electroweak mass eigenstates
\be
W^\pm = \frac1{\sqrt{2}}(W^1 \pm i W^2)\ , \quad \label{eq:eweigen}
\left( \begin{array}{c}
             Z\\
             A 
             \end{array}\right) = 
             \left( \begin{array}{cc}
             c_w & -s_w \\
             s_w & c_w
             \end{array}\right)
 \left( \begin{array}{c}
              W^3\\
               B
             \end{array}\right) \  ,
\ee
we obtain the following couplings:
\bea
\label{eq:stensor}
\Gamma^{\mu\nu}_{SV_1V_2}&=& \frac{g_{sV_1V_2}}{m_S}  (p_{V_1}\cdot p_{V_2} g^{\mu\nu} -p_{V_1}^\nu p_{V_2}^\mu)  \ , \\
 \label{eq:singscoup}
g_{Sgg}&=& \kappa_{g} \frac{\alpha_{s}}{4\pi}\ , \\
g_{SWW}& =& \kappa_{W} \frac{\alpha_{em}}{4\pi s_w^2}\ ,\\
 g_{SZZ}&=& \frac{\alpha_{em}}{4\pi}\left(\kappa_W \frac{c_w^2}{s_w^2} +\kappa_B \frac{s_w^2}{c_w^2}\right)\ , \\
g_{SZ\gamma}&=& \frac{\alpha_{em}}{4\pi} c_w s_w \left(\frac{\kappa_W}{s_w^2}-\frac{\kappa_B}{c_w^2}\right) \ ,\\
g_{S\gamma\gamma} &=& \frac{\alpha_{em}}{4\pi} (\kappa_W+\kappa_B)\  .
\eea
The partial decay widths into $V_1V_2$ are given by the following expressions, 
which are valid even if the massive gauge bosons are off-shell:
\bea
\label{eq:sgg}
\Gamma (S\to gg)
&=& \frac{1}{8\pi} \, \vert {g}_{Sgg} \vert^2
m_S \ ,\\
\label{eq:sgaga}
\Gamma (S\to \gamma\gamma)
&=& \frac{1}{64\pi} \, \vert {g}_{s\gamma\gamma} \vert^2\,
m_S \ , \\
\Gamma (S\to Z\gamma)
&=&\int_0^{m_S^2} dm_1^2 \, \frac{1}{32\pi} \, \vert {g}_{SZ\gamma} \vert^2\,
m_S\, \left(1-\frac{m_1^2}{m_S^2}\right)^3 \,P_1 \ ,
\label{eqn:offshellZgamma} \\
\label{eq:swz}
\Gamma(S\to V_1'V_2')& =& \int_0^{m_S^2} dm_1^2 \int_0^{\left(m_S-\sqrt{m_1^2}\right)^2} 
\hspace*{-10pt}dm_2^2\, \frac{\delta_{V'}\, m_1^3m_2^3}{16\pi\, m_S^5} 
\, |{g}_{SV'_1V'_2}|^2\,\gamma_b (2\gamma_a^2+1)\,   P_1 P_2\ ,
\eea
where $V_1'V_2'=\{W^+W^-,ZZ\}$, and $\delta_{V'}=1$ for $W^+W^-$ and $2$ for $ZZ$. Moreover, 
\be
 \gamma_a = \frac{1}{2m_1m_2}\left[ m_S^2 - (m_1^2+m_2^2) \right] \ , \qquad
\gamma_b = \sqrt{\gamma_a^2 -1}\ ,
\ee
and
\be
P_i = \frac{M_{V_i}\Gamma_{V_i}}{\pi}\frac{1}
{(m_i^2-M_{V_i}^2)^2+M_{V_i}^2\Gamma_{V_i}^2}
\ee
are the propagator factors for $W/Z$ bosons which, in the narrow width approximation for on-shell particles, become just $\delta(m_i^2-M_{V_i}^2)$.

A few comments are in order. 
First, the phase space factor in the $gg$ channel is a factor of 8 larger than 
that in $\gamma\gamma$ because of color. Moreover, when the singlet scalar is light, 
below the kinematic threshold of $WW$ bosons, decays into massive final states are not 
preferred generically because of kinematic suppression. Therefore we see a general pattern of partial widths into $V_1V_2$:
\be
\label{eq:pattern}
\Gamma_{gg}\  \agt\  \Gamma_{\gamma\gamma} \ \agt\  \Gamma_{Z\gamma}\  \agt \ \Gamma_{WW} \ \agt\  \Gamma_{ZZ} 
\ , \quad {\rm for}  \quad m_S \alt 2m_W \ ,
\ee
which is in sharp contrast with that of the Higgs boson for which $\Gamma_{WW}$ dominates even below the $WW$ threshold. 
A singlet scalar prefers to decay into $gg$, $\gamma\gamma$, and to a lesser extent $Z\gamma$. 

Next, the gluonic coupling $g_{Sgg}$ is responsible for producing the singlet scalar in the gluon fusion channel, 
which is also the dominant production channel of the Higgs in hadron colliders. For a SM Higgs the gluonic coupling is 
\be
 \frac23 \frac{\alpha_s}{2\pi} \frac{h}{4v} G_{\mu\nu}^a G^{a\,\mu\nu} \ ,
\ee
which is related to the top quark contribution to the one-loop beta function of QCD via
 the Higgs low-energy theorem \cite{Ellis:1975ap,Shifman:1979eb}.  In fact, the 
 factor of $2/3$ is exactly $b_F^{(3)}$, the contribution to the QCD running from a Dirac fermion. 
 When there exist a pair of heavy vector-like fermions $(Q^c, Q)$ in the fundamental 
 representation of $SU(3)_c$ coupling to $S$ with the interactions
\be
m_Q Q^c Q + y_Q S\, Q^c Q\ ,
\ee
it induces the gluonic coupling \cite{Low:2009di}:
\be
g_{Sgg} = \frac{\alpha_s}{3\pi} {m_S \over m_Q} y_Q.
\ee
Strictly speaking, the low-energy theorem applies only when the mass of the particle in the loop is 
much larger than the scalar mass,
$m_S^2/(4m_Q^2)\ll 1$, so that
the loop diagram can be approximated by a dimension five operator. In practice it is found that the 
effective operator is an excellent approximation even when the Higgs mass is as heavy as 1 TeV, 
although the top mass is only 172 GeV$/c^2$ \cite{Spira:1995rr}. 
In this case we see the ratio of the production rates
\be
r_g \equiv  \frac{\sigma(gg\to S)}{\sigma(gg\to h)_{SM}} = \frac{v^2}{m_Q^2} \ .
\ee
Two benchmark scenarios are: 1) $m_Q\sim 250$ GeV$/c^2$
and $r_g \simeq 1$ and 2) $m_Q \sim 750$ GeV$/c^2$ and $r_g\sim $ 10\%.\footnote{Depending on how the $m_Q\sim 250$ GeV$/c^2$
quark couples to SM quarks, it may or may not be excluded by Tevatron searches.}

The final comment concerns the branching fraction of the diphoton mode. 
We will make the assumption that widths of $S$ decaying into fermion pairs 
are much smaller than those into the vector boson pairs, which seems plausible 
given the singlet nature, and thus the total width of $S$ can be approximated by 
the widths into $V_1V_2$ alone. The branching fraction of the $\gamma\gamma$ channel 
is then largely determined by the relative magnitude of $\Gamma_g$ versus $\Gamma_\gamma$, which is
\be
\frac{\Gamma_\gamma}{\Gamma_g} = \frac18 \left(\frac{g_{S\gamma\gamma}}{g_{Sgg}}\right)^2  \ .
\ee
Naively the above ratio is in the order of $\alpha_{em}^2/(8\alpha_s^2)\sim 6\times 10^{-4}$. 
In reality there are strong enhancements in $g_{S\gamma\gamma}$ due to various reasons. 
For example,  in the SM $g_{h\gamma\gamma}$ receives contributions from the $W$ boson 
as well as the top quark loops, which in the limit of $M_W, m_t \to \infty$ are\footnote{In the SM the $m_W\to \infty$ limit is not valid; we are interested in this limit only in the context of understanding the coupling of $S$.}
\be
\label{eq:SMcoupling}
g_{h\gamma\gamma}^{(W)}= -7  \times \frac{\alpha_{em}}{2\pi}\frac{m_S}{v} \ , \qquad g_{h\gamma\gamma}^{(t)}= \frac43 N_c \, e_t^2 \times \frac{\alpha_{em}}{2\pi}\frac{m_S}{v} =  \frac{16}9 \times \frac{\alpha_{em}}{2\pi} \frac{m_S}{v}\ ,
\ee
where $-7$ and $4/3$ are the contribution to the QED running from the $W^\pm$ boson and the top quark, respectively, while $N_c$ is the number of colors and $e_t$ is the electric charge of the top quark.\footnote{Here we see explicitly that the top quark, or any fourth generation fermion, interferes destructively with the $W$ boson loop.} Thus in the SM the ratio of partial widths into $\gamma\gamma$ and $gg$ is enhanced from $\alpha_{em}^2/(8\alpha_s^2)$ to
\be
\frac{\Gamma^{(SM)}_\gamma}{\Gamma^{(SM)}_g} \sim 0.04 \ .
\ee
This demonstrates that if $\kappa_W$ and $\kappa_B$ are induced by a set of heavy electroweak gauge bosons \cite{Cao:2009ah}, or three new colored fermions, we could easily have
\be 
 \frac{\Gamma_\gamma}{\Gamma_g} \sim {\cal O}(0.05)  \ ,
 \ee
 which, under our earlier assumption on the total width, in turns implies the branching fraction into two photons is enhanced from that of the SM Higgs by
 \be
 \frac{Br(S\to \gamma\gamma)}{Br(h\to\gamma\gamma)} \sim \frac{0.05}{2\times 10^{-3}} \sim {\cal O}(20{\rm -}30) \ .
 \ee 
It is worth emphasizing that such a large enhancement does not require a large multiplicity of new particles. In terms of $\kappa_W$ and $\kappa_B$, the corresponding values for the SM couplings in Eq.~(\ref{eq:SMcoupling}) are
\be
(\kappa_W + \kappa_B)^{(W)}= -14 \frac{m_S}{v}\ , \qquad (\kappa_W + \kappa_B)^{(t)}= \frac{32}9\, \frac{m_S}{v} \ .
\ee
Thus we see that both of them could easily be in the order ${\cal O}(1-10)$.
These large values of the couplings imply relatively low masses for the exotics propagating in the
loops that generate them, but this need not conflict with experimental bounds if, for instance, the
exotics carry a new quantum number requiring them to be produced in pairs.

\begin{figure}
\includegraphics[scale=0.45, angle=0]{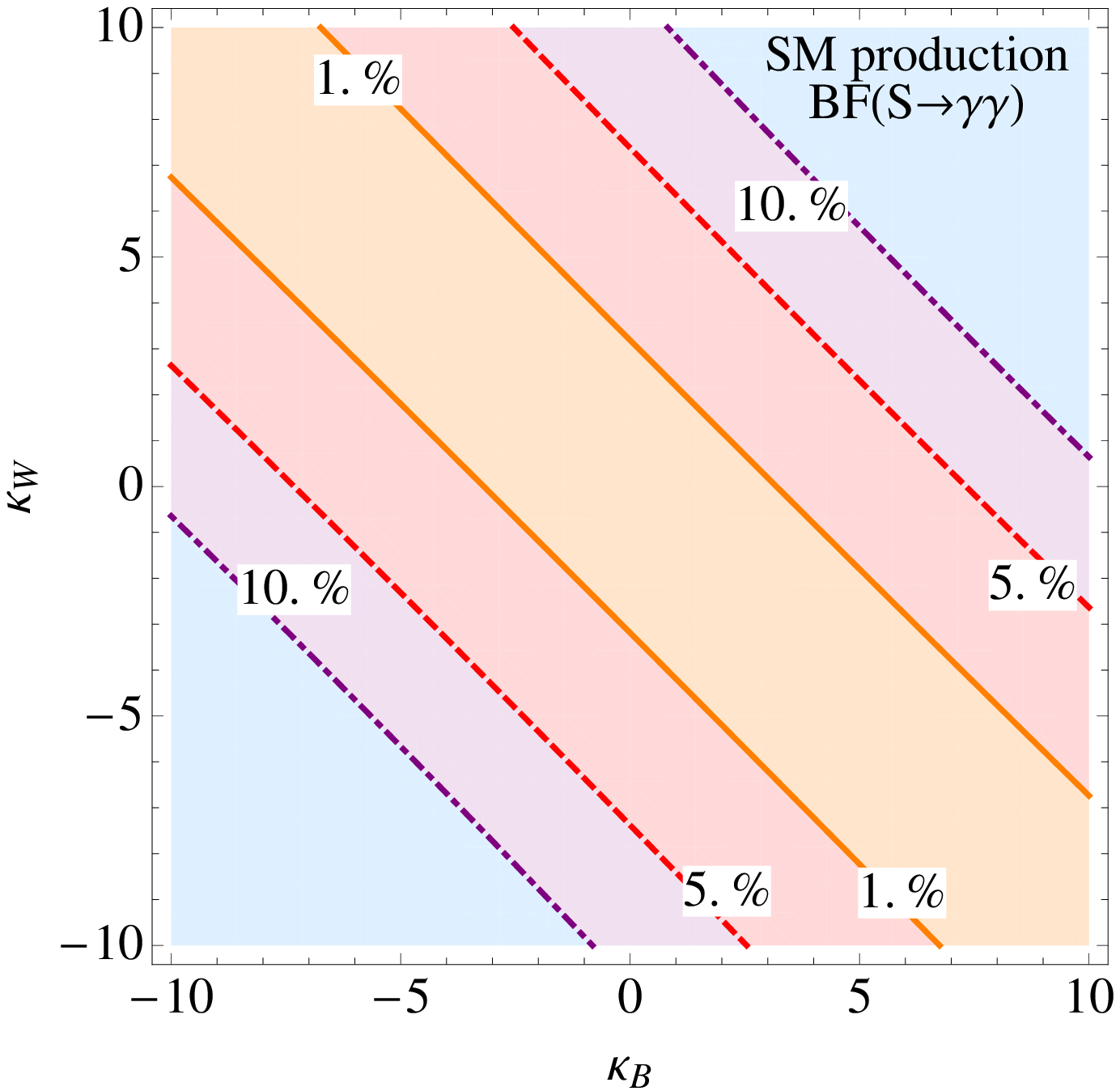}
\includegraphics[scale=0.45, angle=0]{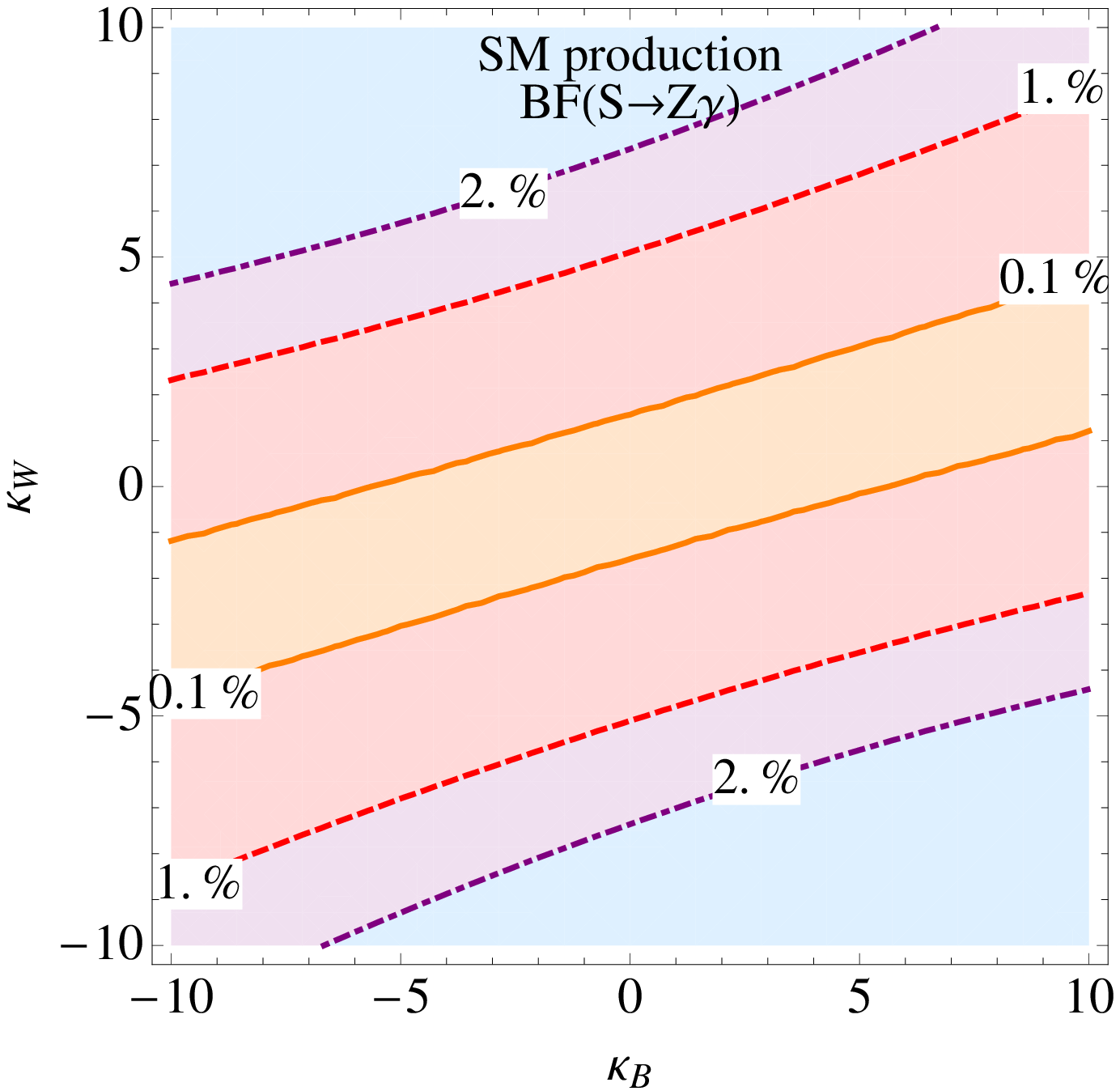}\\
\includegraphics[scale=0.45, angle=0]{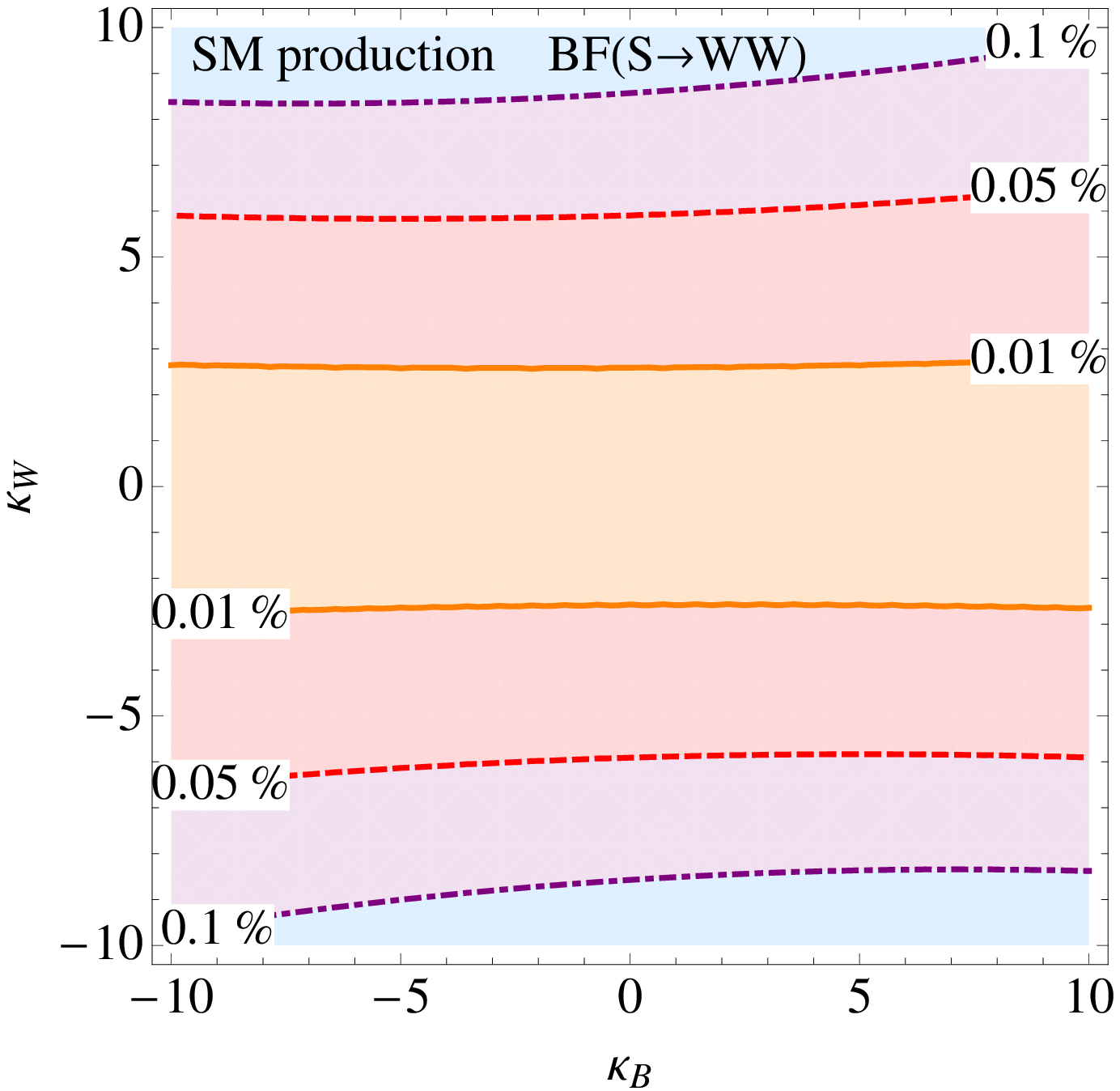}
\includegraphics[scale=0.45, angle=0]{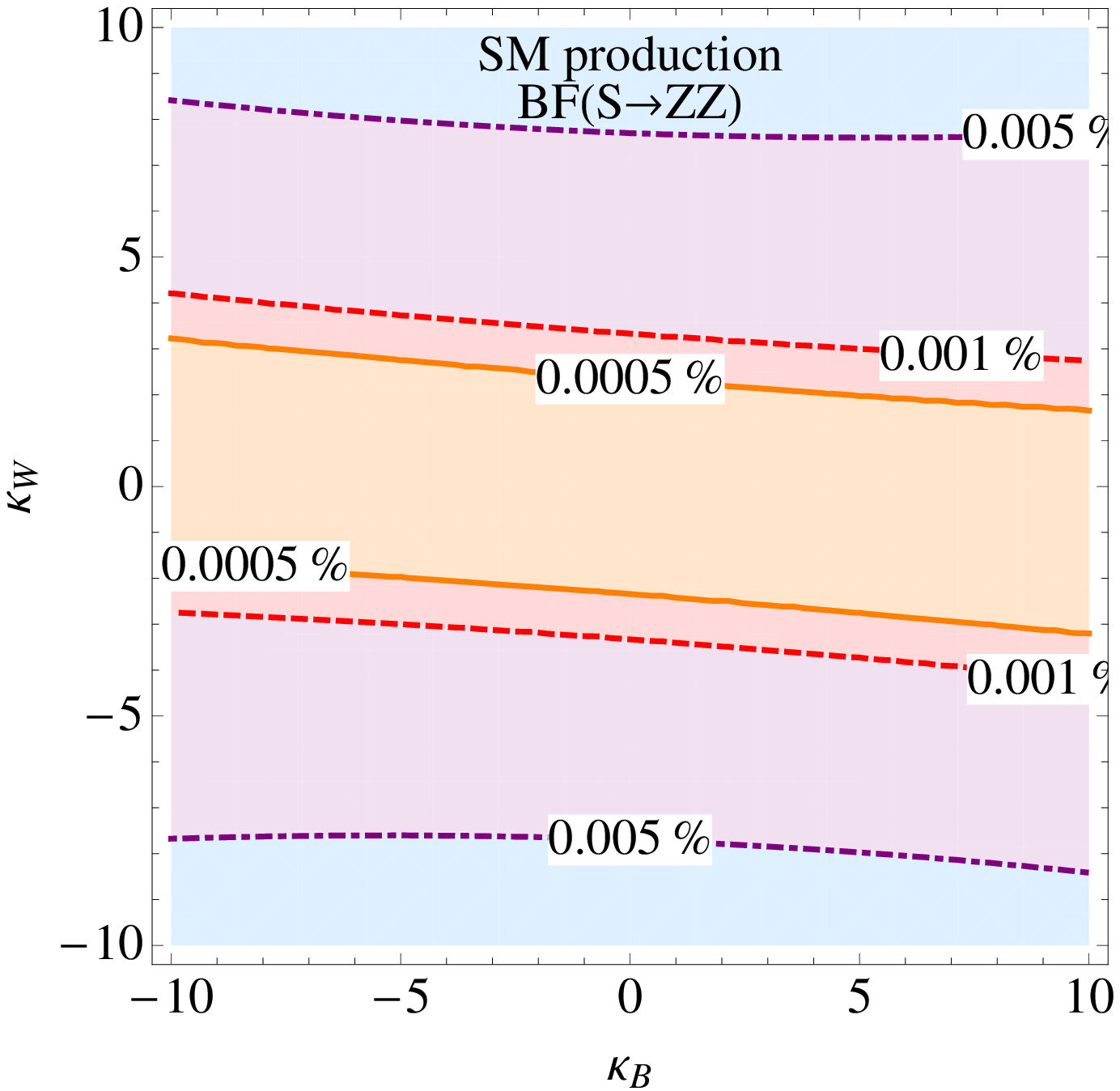}
\caption{{\em Decay branching fractions of a singlet scalar $S$ with $m_S=115$ GeV$/c^2$ into pairs of 
electroweak vector bosons. In the plot we assume the production rate of $S$ is the same as that of a SM 
Higgs with the same mass. In general the $\gamma\gamma$ mode has the largest branching fraction, 
followed by $Z\gamma$, $WW$, and $ZZ$ channels.}
}
\label{fig:BF}
\end{figure}

In Fig.~\ref{fig:BF} we show the branching fractions, as functions of $\kappa_W$ and $\kappa_B$, of an electroweak singlet scalar with a mass of 115 GeV$/c^2$ 
decaying into all four pairs of electroweak gauge bosons, assuming the same gluonic partial width as in the SM. 
Indeed we see the pattern in Eq.~(\ref{eq:pattern}), as expected on general grounds, holds up very well.

\section{Phenomenology}\label{sec:pheno}

\begin{figure}[b]
\includegraphics[scale=0.45, angle=0]{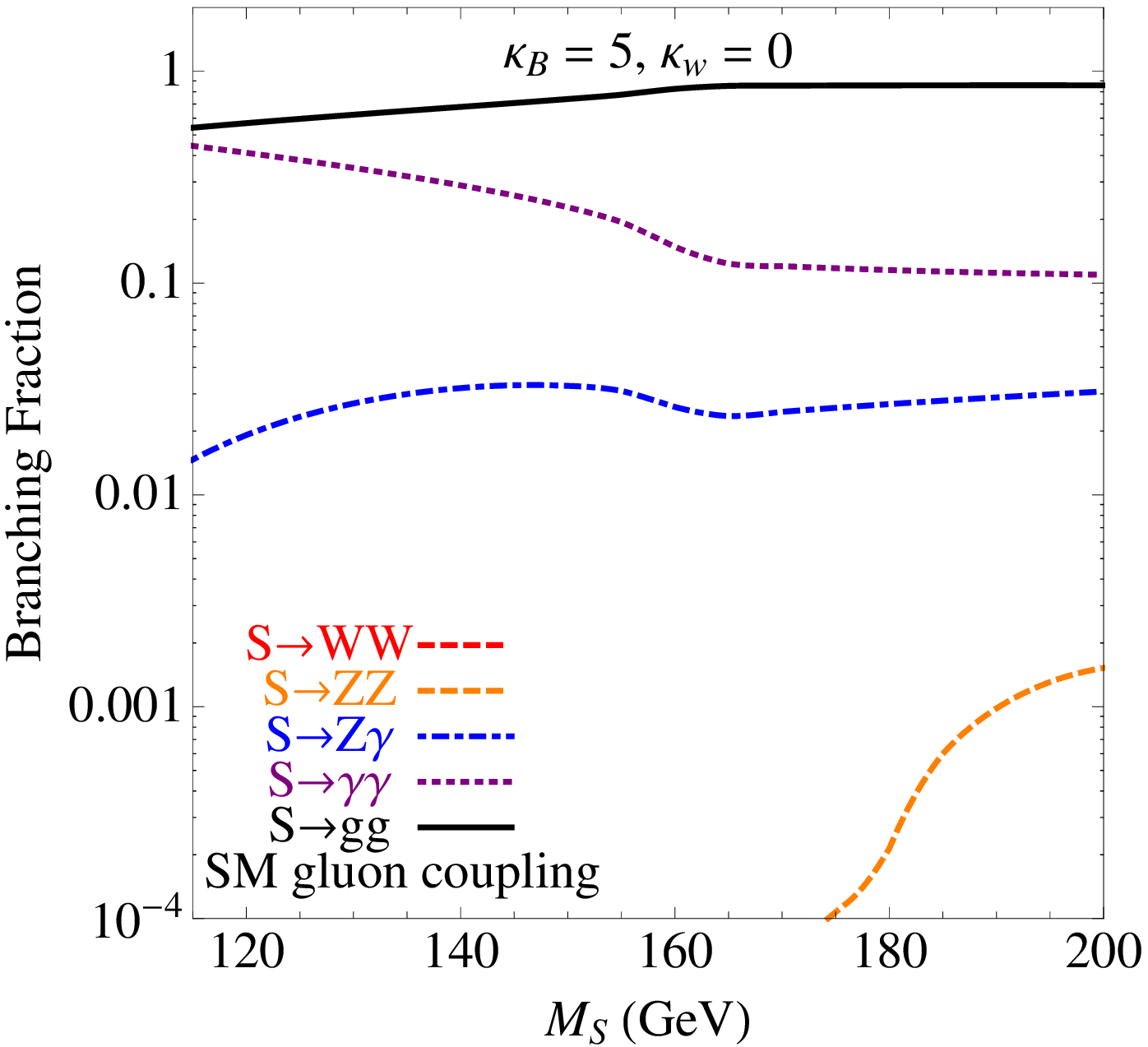}
\includegraphics[scale=0.45, angle=0]{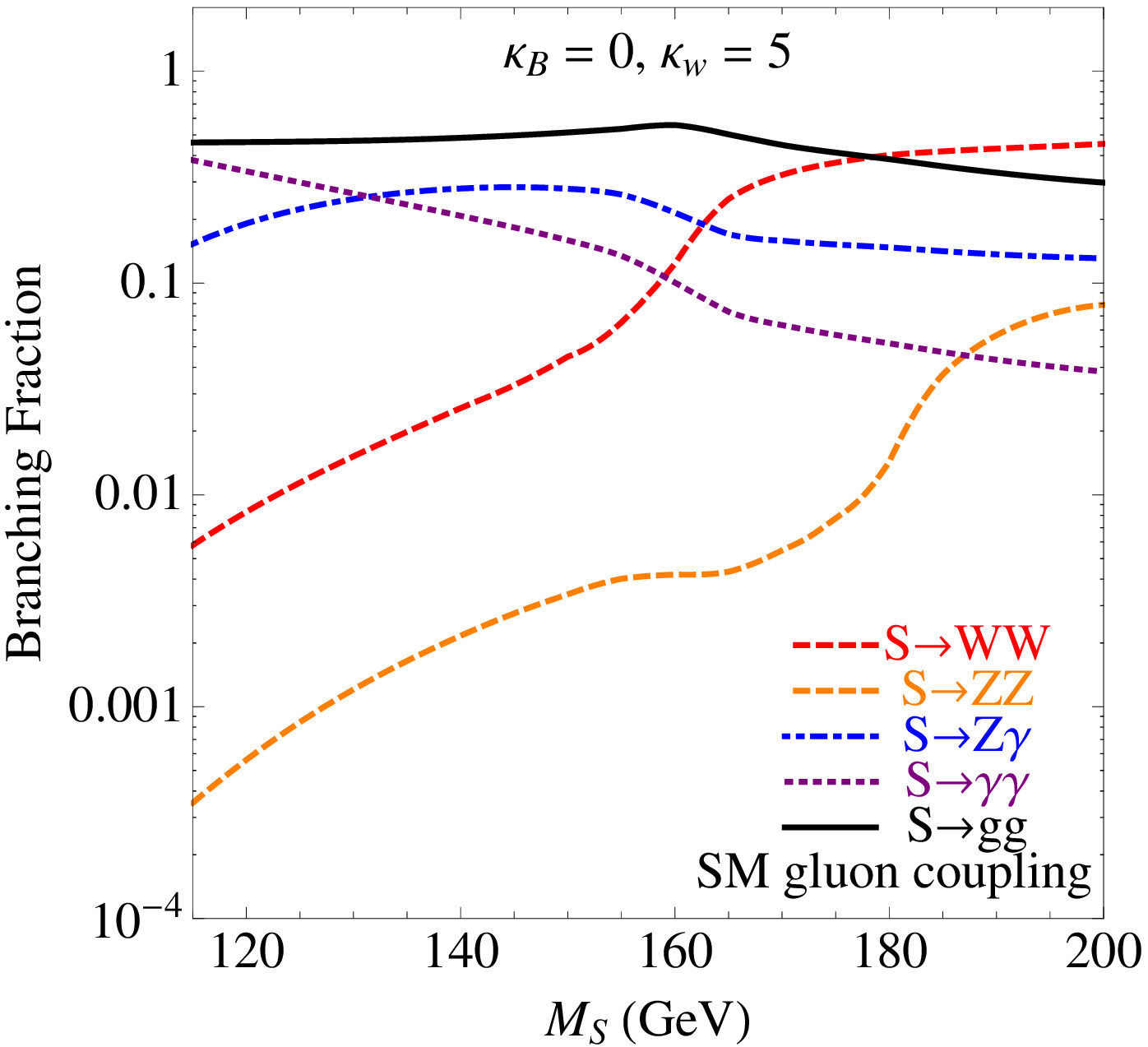}\\
\includegraphics[scale=0.45, angle=0]{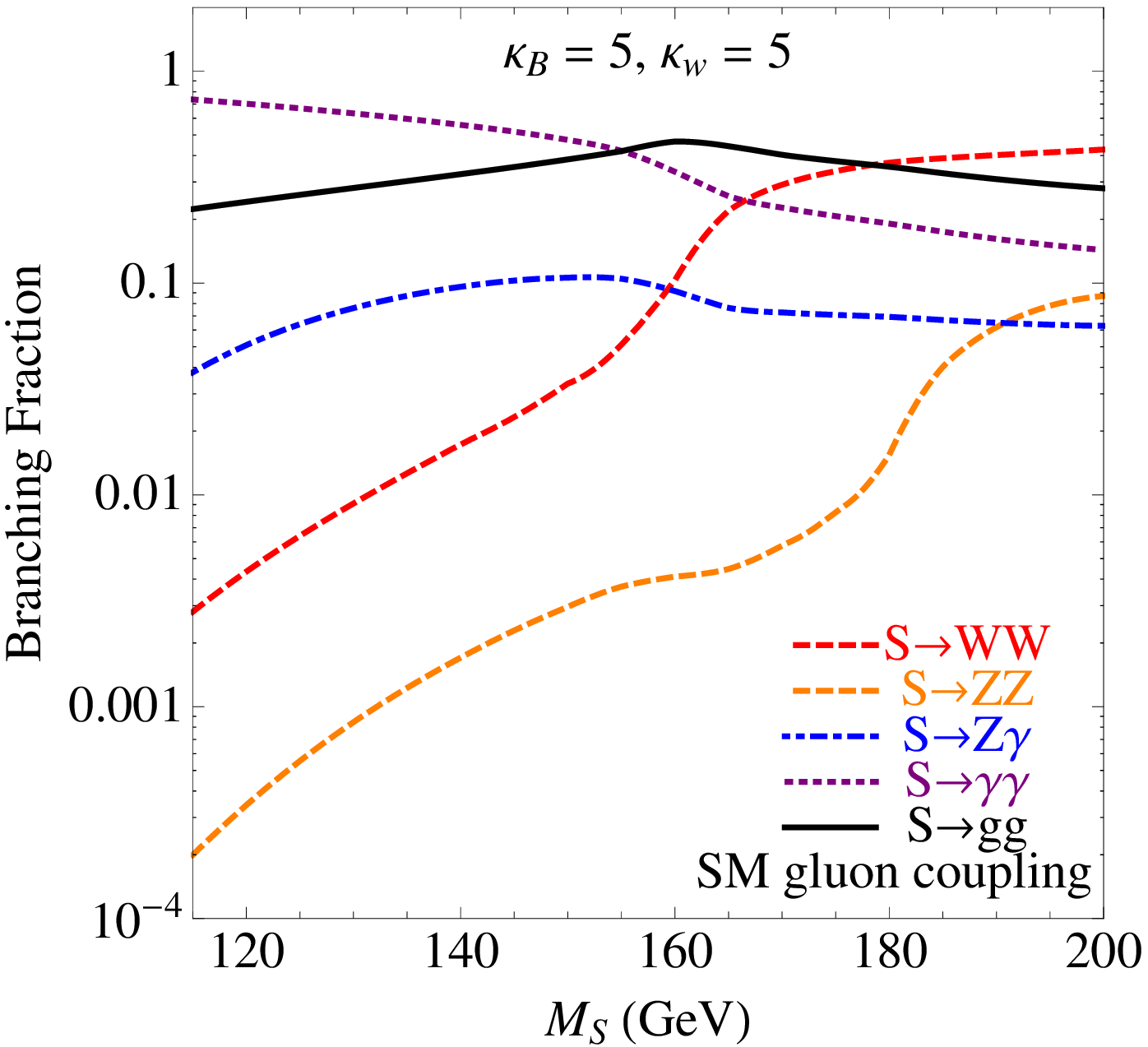}
\includegraphics[scale=0.45, angle=0]{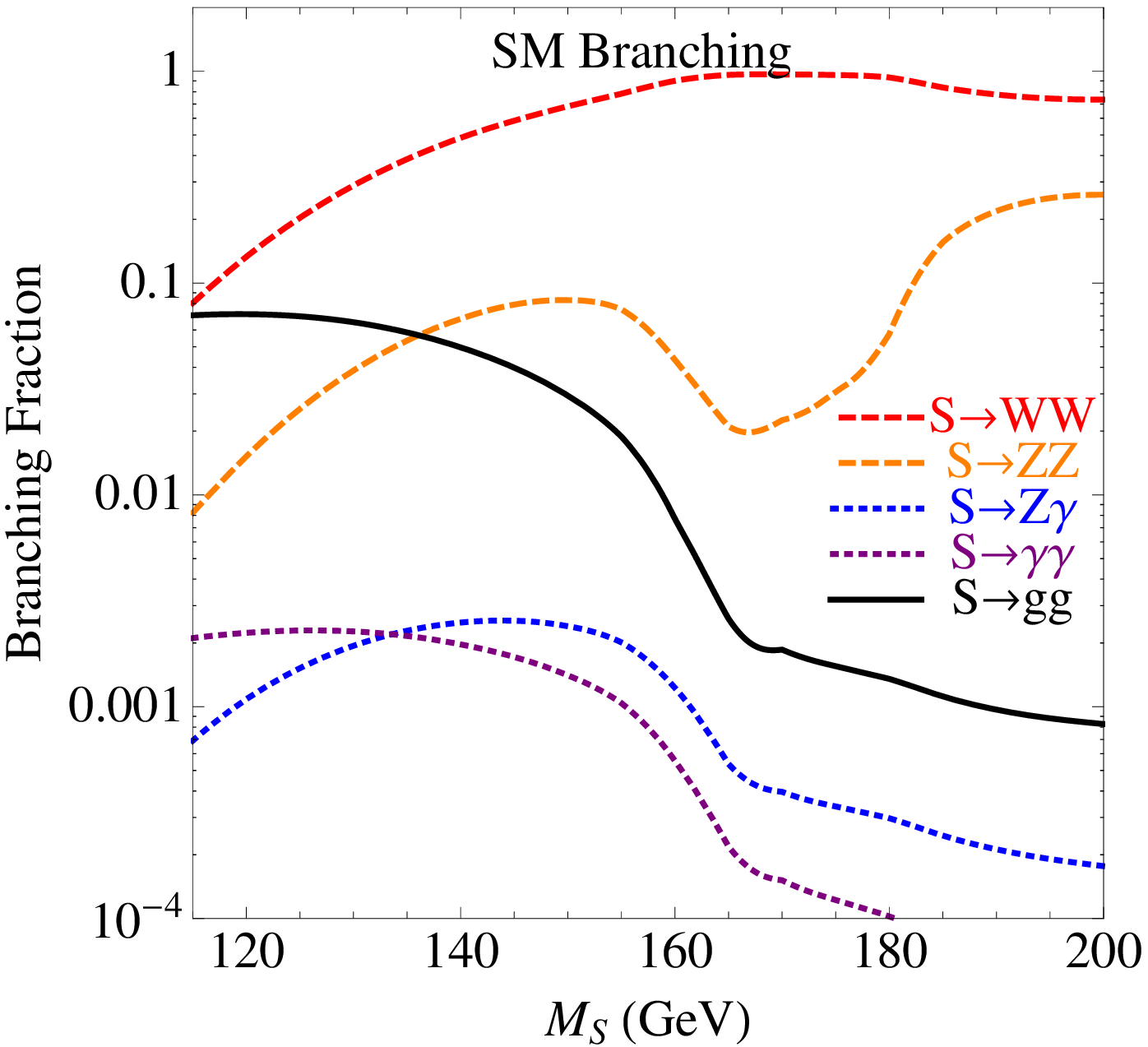}
\caption{{\em Decay branching fractions of a singlet scalar $S$ into pairs vector bosons as a function of mass, for three different choices of $\kappa_W$ and $\kappa_B$, assuming a SM coupling strength to gluons.  For comparison  the branching fractions for a SM Higgs boson is also shown in the lower-right figure.}
}
\label{fig:BFmscan}
\end{figure}

\begin{figure}[b]
\includegraphics[scale=0.45, angle=0]{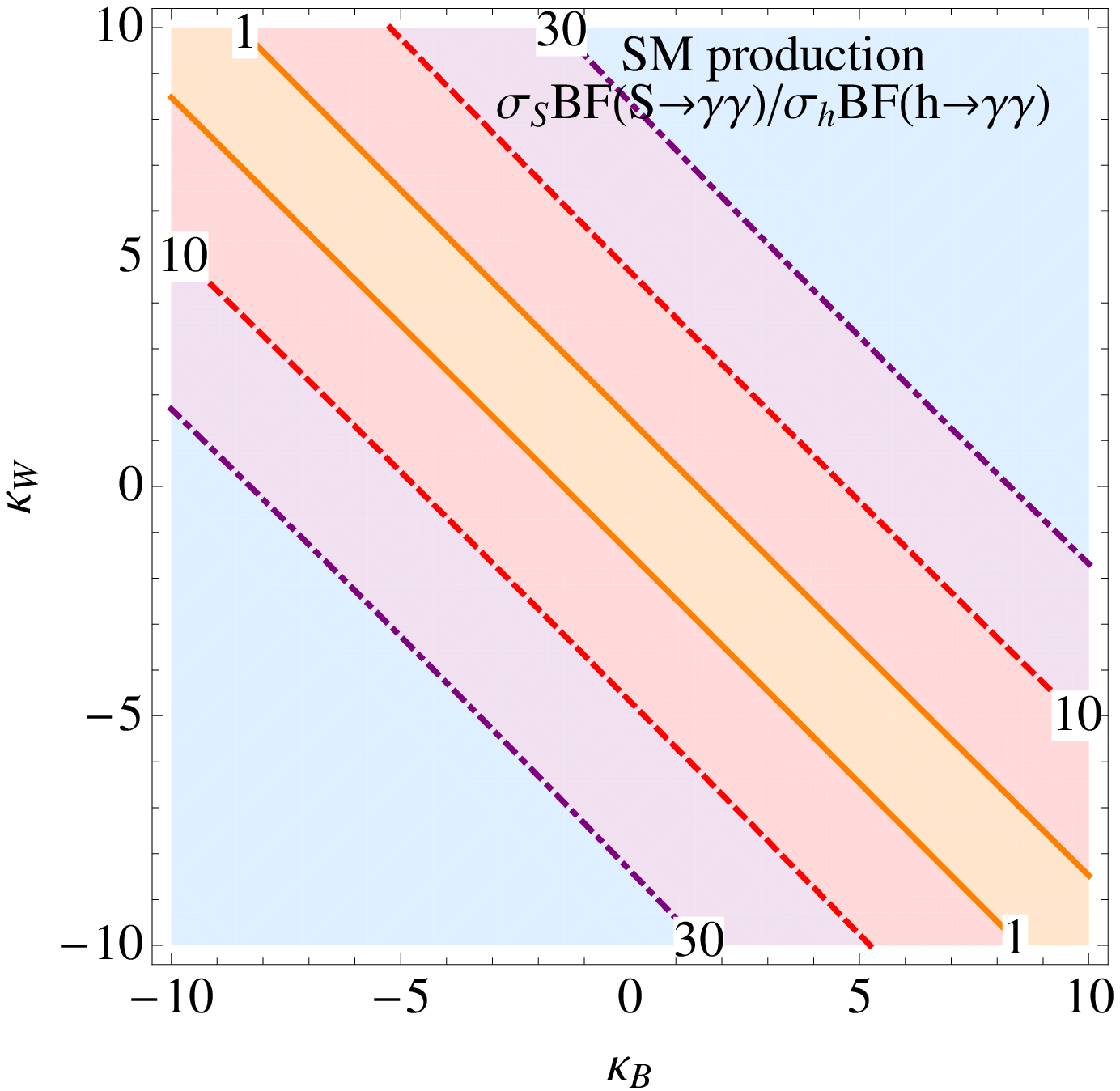} 
\includegraphics[scale=0.45, angle=0]{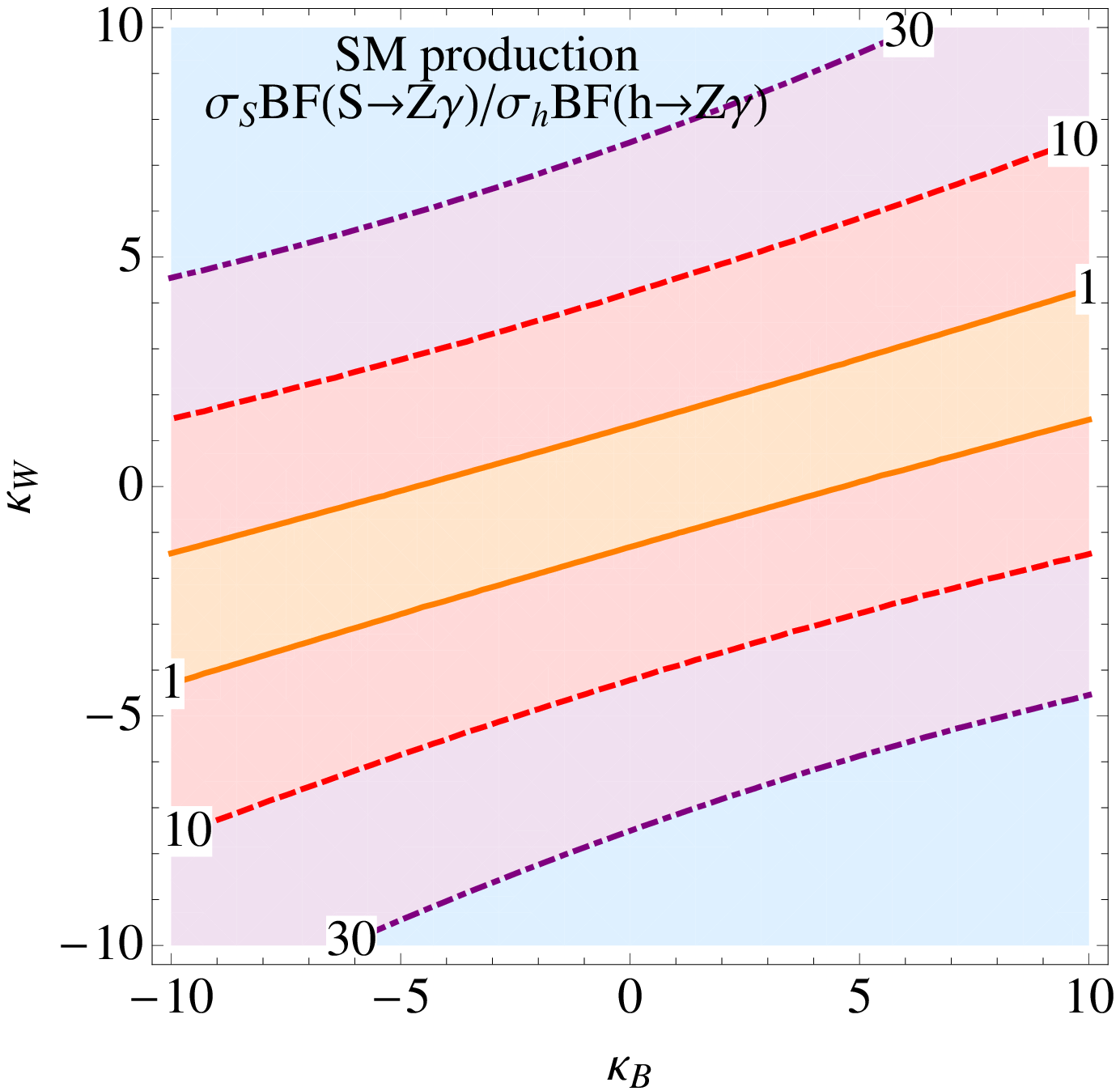}\\
\includegraphics[scale=0.45, angle=0]{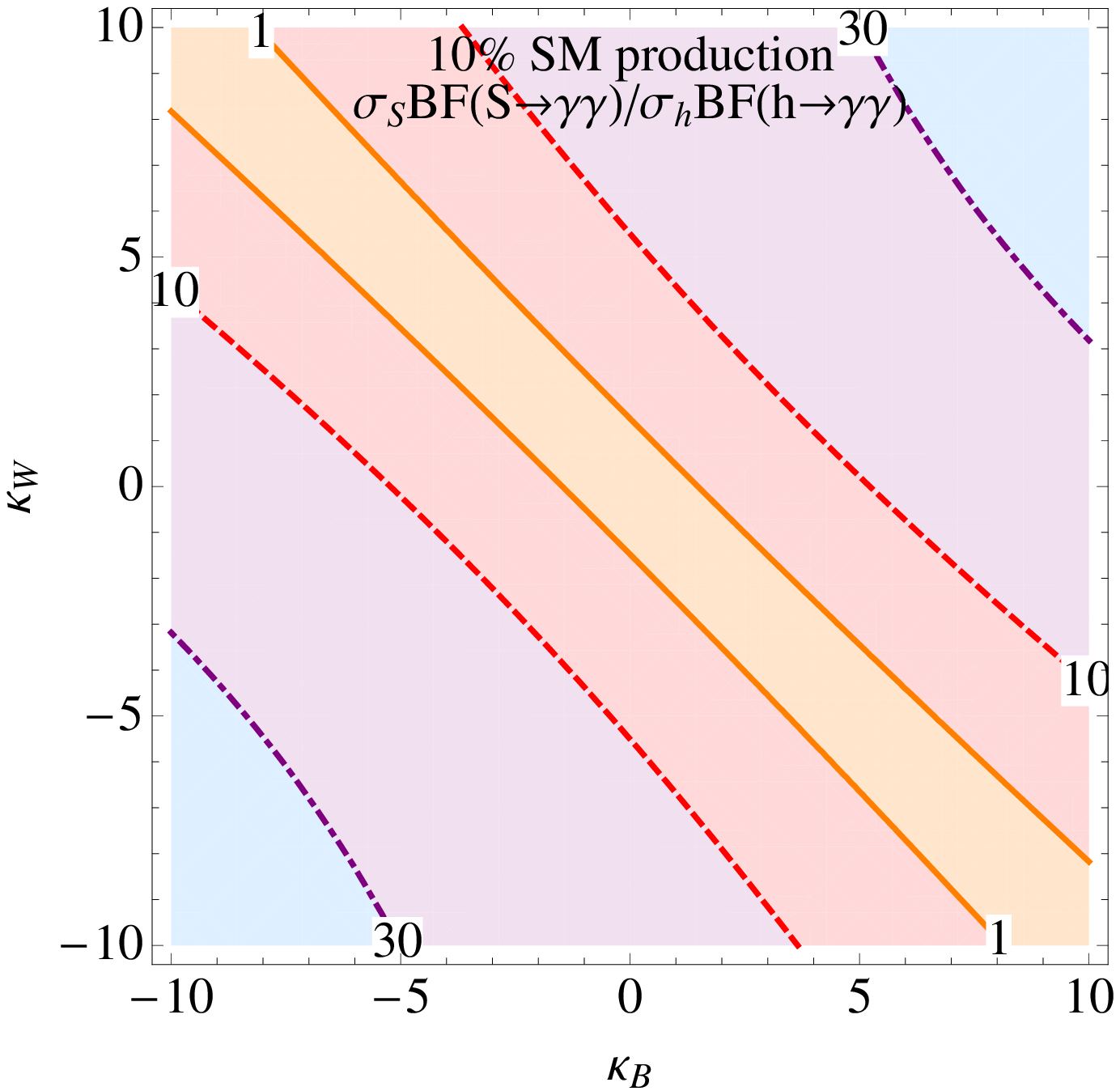} 
\includegraphics[scale=0.45, angle=0]{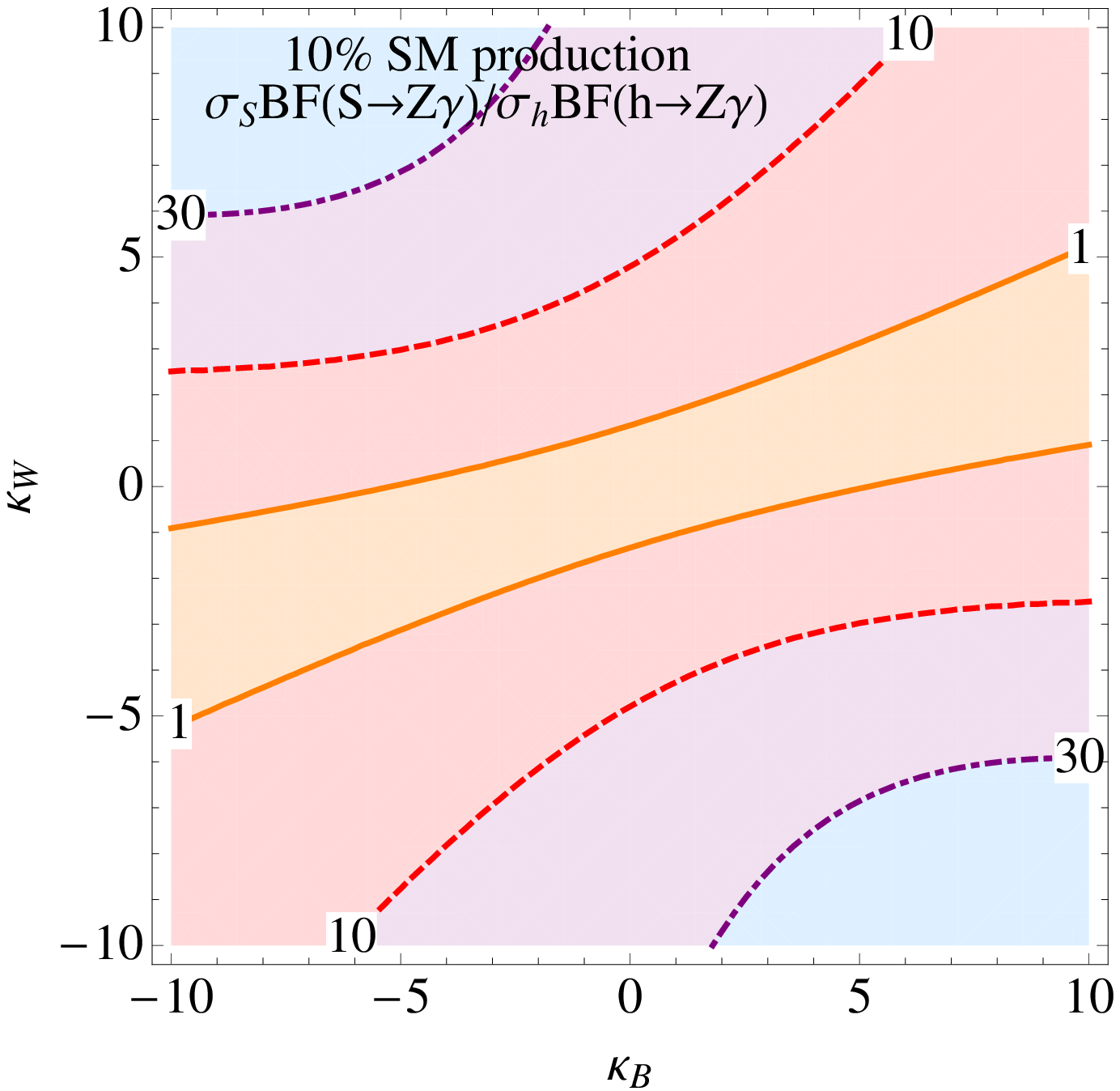}
\caption{\label{fig1}{\em Enhancement of $B\sigma(gg\to S\to \gamma\gamma)$ and $B\sigma(gg\to S\to Z\gamma)$ relative to the SM for $m_S=115$ GeV$/c^2$. Top two plots assume the same production rate as in the SM while the bottom plots assume 10\% of the SM production. Enhancements of ${\cal O}(10 - 30)$ are seen for both the $\gamma\gamma$ and $Z\gamma$ modes.}
}
\label{fig:gamgam}
\end{figure}
We first study the mass dependence of the branching fractions into all five pairs of gauge bosons. In Fig.~\ref{fig:BFmscan} 
we show the decay branching fractions as a function of the scalar mass up to 200 GeV$/c^2$, as well as the corresponding branching fractions 
for a SM Higgs boson, for three different choices of $\kappa_W$ and $\kappa_B$. In the first case, $\kappa_W=0$ and $\kappa_B=5$, the decay into $WW$ is completely absent and the $\gamma\gamma$ mode is the most promising channel for discovery. Such a scenario occurs when the heavy particles mediating the loop-induced couplings between the singlet scalar and the vector bosons are
themselves $SU(2)_L$ singlets. In the opposite scenario when the mediating particles do not carry $U(1)_Y$ charge, 
we consider $\kappa_W=5$ and $\kappa_B=0$. For $m_S$ above the $WW$ kinematic threshold, the $WW$ branching 
fraction dominates over $\gamma\gamma$ and $Z\gamma$, as has been noted previously,
but still is a factor of 3 smaller than that of a SM Higgs. In addition, in this case the $Z\gamma$ mode has a larger branching fraction 
than the $\gamma\gamma$ mode. Nevertheless, after taking into account the decay branching ratio of a $W/Z$ boson into $\ell\nu/\ell\bar{\ell}$ final states,  we see that the 
$\gamma\gamma$ and $Z\gamma$ modes are still competitive with $WW$ mode as the discovery channels. 
In the last possibility we consider, $\kappa_W=\kappa_B=5$, the diphoton mode has a branching fraction larger than $Z\gamma$ 
and a few times smaller than $WW$. Again after including decays into leptonic final states for $W$ and $Z$ bosons, the diphoton channel remains the most likely channel for discovery.

It is worth pointing out that recently Tevatron presented exclusion limits for
a SM Higgs boson in the mass region between 158 and 172 GeV$/c^2$ at 95\% confidence level \cite{Aaltonen:2011gs}. 
The most stringent limit comes at 165 GeV$/c^2$ where the event rate of a Higgs decaying into the $WW$ channel 
is constrained to be less than roughly 65\% of the SM event rate. In the cases presented in Fig.~\ref{fig:BFmscan} 
we see the decay branching fraction of a singlet scalar into $WW$ is about 10 -- 20 \% at the mass of 165 GeV$/c^2$, 
while that of the SM Higgs is nearly 90\%. Thus a singlet scalar can easily escape constraints from the Tevatron limit 
even if it has a comparable production cross section to the SM Higgs.
\begin{table}[b]
\caption{Production cross sections of the $\gamma\gamma$ and $Z\gamma$ channels through the SM Higgs boson at 7 TeV and 14 TeV LHC energies. }
\begin{center}
\begin{tabular}{|c|cc|cc|}
\hline
& $B\sigma_{\gamma\gamma}$ at $\sqrt s = 7$ TeV & $B\sigma_{\gamma\gamma}$ at $\sqrt s = 14$ TeV\\
\hline
$m_h=115$ GeV$/c^2$  & 26 pb & 85 fb\\
$m_h=130$ GeV$/c^2$  & 21 pb& 72 fb\\
$m_h=145$ GeV$/c^2$  & 12 pb& 44 fb\\
$m_h=165$ GeV$/c^2$  & 1.2 pb& 4.3 fb\\
$m_h=180$ GeV$/c^2$  & 0.44 pb& 1.7 fb\\
$m_h=200$ GeV$/c^2$  & 0.18 pb& 0.7 fb\\
\hline 
\hline
 & $B\sigma_{Z\gamma}$ at $\sqrt s = 7$ TeV &  $B\sigma_{Z\gamma}$ at $\sqrt s = 14$ TeV\\
\hline
$m_h=115$ GeV$/c^2$  & 8.5 pb & 28 fb\\
$m_h=130$ GeV$/c^2$  & 18 pb& 62 fb\\
$m_h=145$ GeV$/c^2$  & 19 pb& 65 fb\\
$m_h=165$ GeV$/c^2$  & 2.9 pb& 11 fb\\
$m_h=180$ GeV$/c^2$  & 1.3 pb& 4.9 fb\\
$m_h=200$ GeV$/c^2$  & 0.59 pb& 2.3 fb\\
\hline 
\end{tabular}
\end{center}
\label{tab:smhprod}
\end{table}%

In what follows we focus on the $\gamma\gamma$ and $Z\gamma$ modes as the most likely discovery channels for an electroweak singlet scalar.
To emphasize the possibility of an early discovery, we show in Fig.~\ref{fig:gamgam} the enhancements of the event rates, $B\sigma=\sigma\times Br$, of the singlet scalar decaying into $\gamma\gamma$ and $Z\gamma$ over the SM expectations.   An increase in the order of 10 -- 30 is common in the region of parameter space we consider. For reference, we provide the production cross sections of these modes obtained in MCFM~\cite{MCFM1,MCFM2} and HDECAY~\cite{HDECAY} through the SM Higgs boson in Table~\ref{tab:smhprod}.  Although not shown in the figure, we point out that suppressions of the event rate in the $WW$ and $ZZ$ channels are on the order of $10^{-2}$ to $10^{-3}$ relative to the SM.

\subsection{$S\to \gamma \gamma$}


Current searches at the Tevatron for the Higgs boson in the $\gamma \gamma$ mode places the limit at ${\cal O}(10-30)$ times the SM cross section of $\sigma(gg\to h\to \gamma \gamma)$, depending on the mass.  The 95\% C.L. exclusion limits from CDF with 7.0 fb$^{-1}$ of data are in the range of $10-20$ times the SM expectation, with the exception at $m_h=120$ GeV$/c^2$ where only a cross section of more than 28x can be constrained, which is $2\sigma$ above the expected sensitivity~\cite{cdf-10485}.  However, this feature is not present in the D0 diphoton analysis with even less data.  At 4.2 fb$^{-1}$ of integrated luminosity, D0 constrains the ratio to be at or below 20 in the $m_h=110-130$ GeV$/c^2$ range~\cite{d0-5858}.

Soon the LHC will have competitive limits in the diphoton channel.  ATLAS reports no excess in the 
diphoton channel to date with 132 pb$^{-1}$ of data~\cite{ATLAS-CONF-2011-071}.  It is expected that within 
the first 1 fb$^{-1}$ of data taking, ATLAS and CMS can limit cross sections of ${\cal O}(3-5)$ times that 
expected from the SM Higgs boson in a mass range of 110-140 GeV$/c^2$~\cite{ATLAS-CONF-2011-004,CMSNote2010_008}

To determine the reach in the $S\to\gamma \gamma$ mode we analyze events at
 $\sqrt s=7$ and 14 TeV up to 10 fb$^{-1}$ and 100 fb$^{-1}$ of integrated luminosity, respectively.  
 For $pp\to S \to \gamma \gamma$, 
we generate events in MadEvent with an effective $hgg$ and $h\gamma\gamma$ coupling~\cite{Alwall:2007st}.  
The irreducible background to the diphoton signal is the continuum $\gamma \gamma$ process.  
We simulate the reducible backgrounds $\gamma j$, $jj$ with jets faking photons at the rate
$\epsilon_{j\to\gamma} = 1.2 \times 10^{-4}$ \cite{Aad:2009wy}.  
We also include Drell-Yan (DY) $Z\to e^+ e^-$, with electrons faking the photons such that
the DY component makes up roughly 10\% of the $\gamma j$ component in the 
$115\text{ GeV$/c^2$}<M_{\gamma\gamma} < 145\text{ GeV$/c^2$}$ range.  
All backgrounds are generated in ALPGEN~\cite{Mangano:2002ea}.  
We apply k-factors to match the NLO cross sections found in Ref.~\cite{Aad:2009wy}.

\begin{table}[t]
\caption{Signal acceptance for trial scalar masses and cross section after level C1 cuts.}
\begin{center}
\begin{tabular}{|c|cc|}
\hline
Acceptance after C2 cut &$\sqrt s=7$ TeV&$\sqrt s=14$ TeV\\
\hline
$m_S=115$ GeV$/c^2$& 0.40& 0.36 \\
$m_S=130$ GeV$/c^2$& 0.43& 0.39\\
$m_S=165$ GeV$/c^2$& 0.46& 0.42\\
\hline\hline
Cross section after C1 cuts (pb) &$\sqrt s=7$ TeV&$\sqrt s=14$ TeV\\
\hline
$\gamma \gamma$& 3.30 & 6.01\\
$\gamma j$& 1.56 & 3.62 \\
$jj$&0.63 & 1.20 \\
$Z\to e^+e^-$& 1.51& 2.92\\
\hline
Total Background (pb) & 7.0 & 13.8\\
\hline
\end{tabular}
\end{center}
\label{tab:sigbkgaa}
\end{table}%

Following the ATLAS cuts in the $\gamma\gamma$ channel~\cite{ATLAS-CONF-2011-004}, we define `C1' cuts:
\begin{eqnarray}
p_{T,\gamma_1} &>& 40\text{ GeV},\quad  p_{T,\gamma_2} > 25\text{ GeV}\\
|\eta_\gamma|&<&2.47,\quad\quad \Delta R_{\gamma\gamma} > 0.4
\end{eqnarray}
where $\gamma_{1,2}$ are the harder and softer photon, respectively, and where $\eta_\gamma$ is the photon's pseudorapidity and $\Delta R_{\gamma\gamma}$ is the separation in the azimuthal-pseudorapidity plane:
\begin{equation}
\Delta R_{i,j} = \sqrt{ \left( \eta_i - \eta_j \right)^2 + \left( \phi_i - \phi_j \right)^2 } \,.
\end{equation}
In addition to the C1 cuts above, we define the signal region in the $\gamma \gamma$ channel 
as having a window around the trial scalar mass with width 5 GeV, defined as our `C2' cut:
\begin{equation}
|M_{\gamma \gamma} - m_S| < 2.5\text{ GeV$/c^2$}.
\end{equation}

Additionally, we account for photons being lost in the ECAL `crack' at $1.37<\eta<1.52$ in the ATLAS detector.  
The detector resolution effects are accounted for by smearing the final-state energy according to \cite{Aad:2009wy}:
\begin{equation}
\frac{\delta E}{E} = \frac{a}{\sqrt{E/{\rm{GeV}}}} \oplus b \,,
\end{equation}
where $a = 10\%$ and $b = 0.7\%$ for photons and leptons.

After applying the cuts outlined above, we find the signal acceptance and background cross sections outlined in Table~\ref{tab:sigbkgaa}.  We find the dominant backgrounds to be $\gamma \gamma$ and $\gamma j$ after requiring the two photons to reconstruct the scalar mass.  

We determine the significance of the signal over the expected background using 
\begin{equation}
S=2(\sqrt{N_S+N_B}-\sqrt{N_B}),
\end{equation}
where $N_S$ and $N_B$ are the expected number of signal and background events.  Likewise, we calculate the 95\% C.L. under the assumption no signal is seen over the expected background.

\begin{figure}[t]
\begin{center}
\includegraphics[scale=0.5, angle=0]{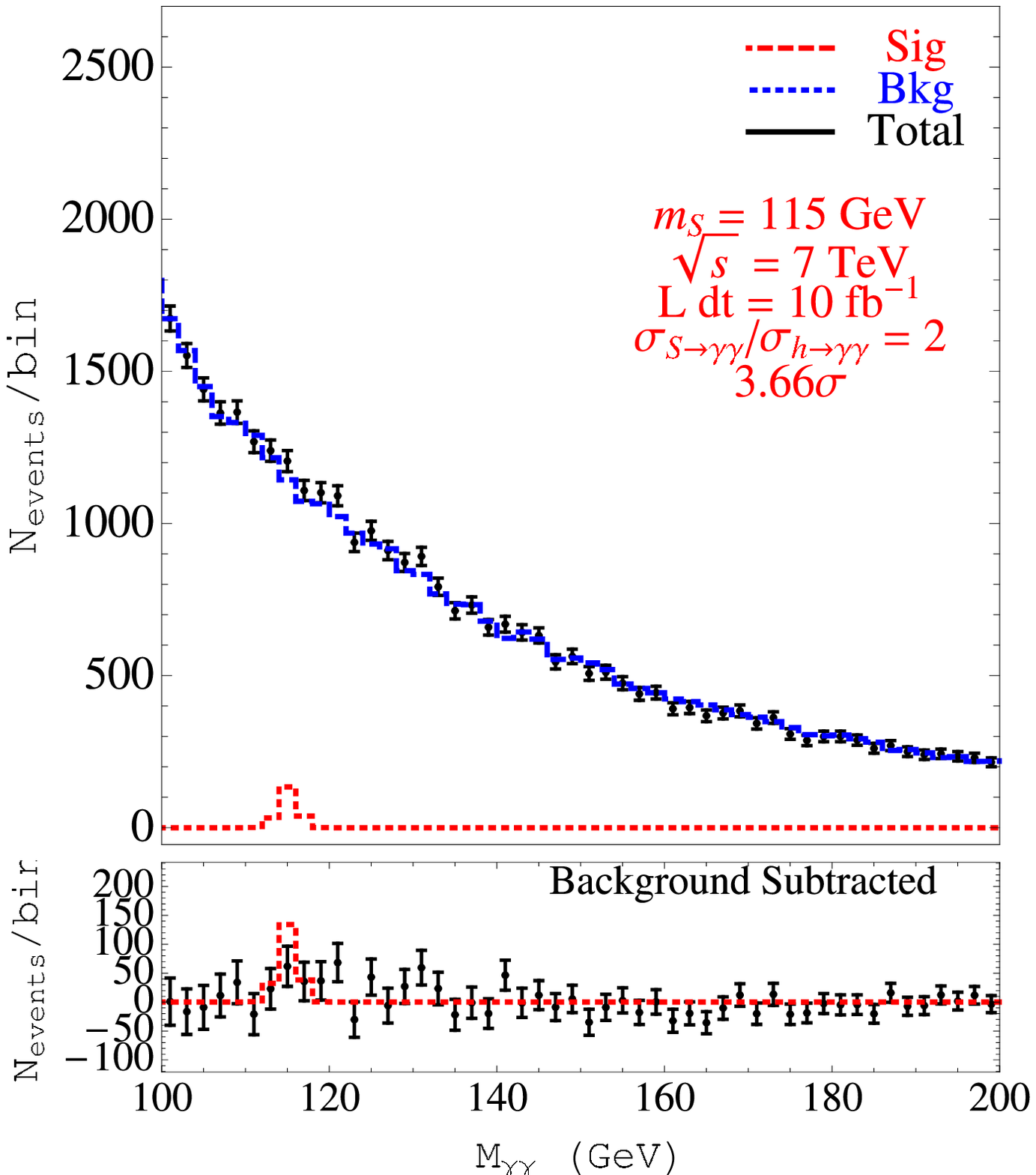}
\includegraphics[scale=0.5, angle=0]{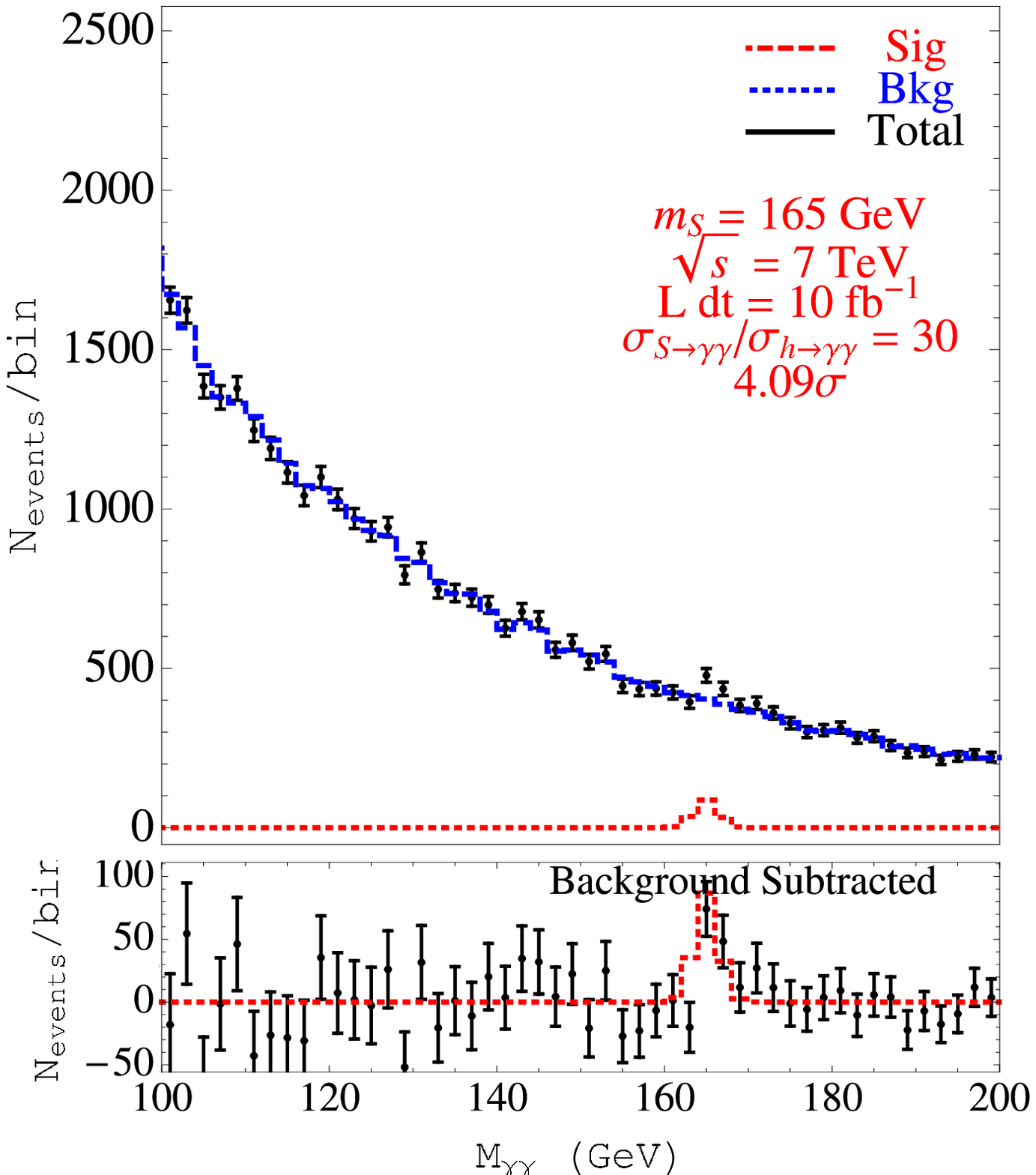}\\\vspace{.1in}
\caption{Distributions of $M_{\gamma\gamma}$ from pseudo-experiments for $M_S=115$ and 165 GeV$/c^2$ and $\sqrt s=7$ TeV for 10 fb$^{-1}$ of integrated luminosity.  The background subtracted 
distribution more clearly shows the possible $\gamma \gamma$ peak.}
\label{fig:s2aadist}
\end{center}
\end{figure}


After applying the above cuts, we find general agreement with the expected sensitivity of the ATLAS and CMS studies with $\sqrt s=14$ TeV and 30 fb$^{-1}$ of integrated luminosity for a Higgs boson at 115 GeV$/c^2$~\cite{atltdr,cmstdr}.  
In Fig.~\ref{fig:s2aadist}, we show the distribution of $M_{\gamma\gamma}$ coming from one pseudo-experiment illustrating 
the prominence of the di-photon peak. 
Of the masses we study, we find the LHC is most sensitive to scalars in the diphoton channel for $m_S=130$ GeV$/c^2$ for 
both 7 and 14 TeV cases.  Due to the opening of the $W^+W^-$ and $ZZ$ channels at higher mass, the expected sensitivity 
of the $\gamma \gamma$ degrades significantly in the SM.  At $m_S=165$ GeV$/c^2$, near the $W^+W^-$ threshold, 
discovery requires about an order of magnitude enhancement in the diphoton signal.  

\begin{figure}[t]
\begin{center}
\includegraphics[scale=0.5, angle=0]{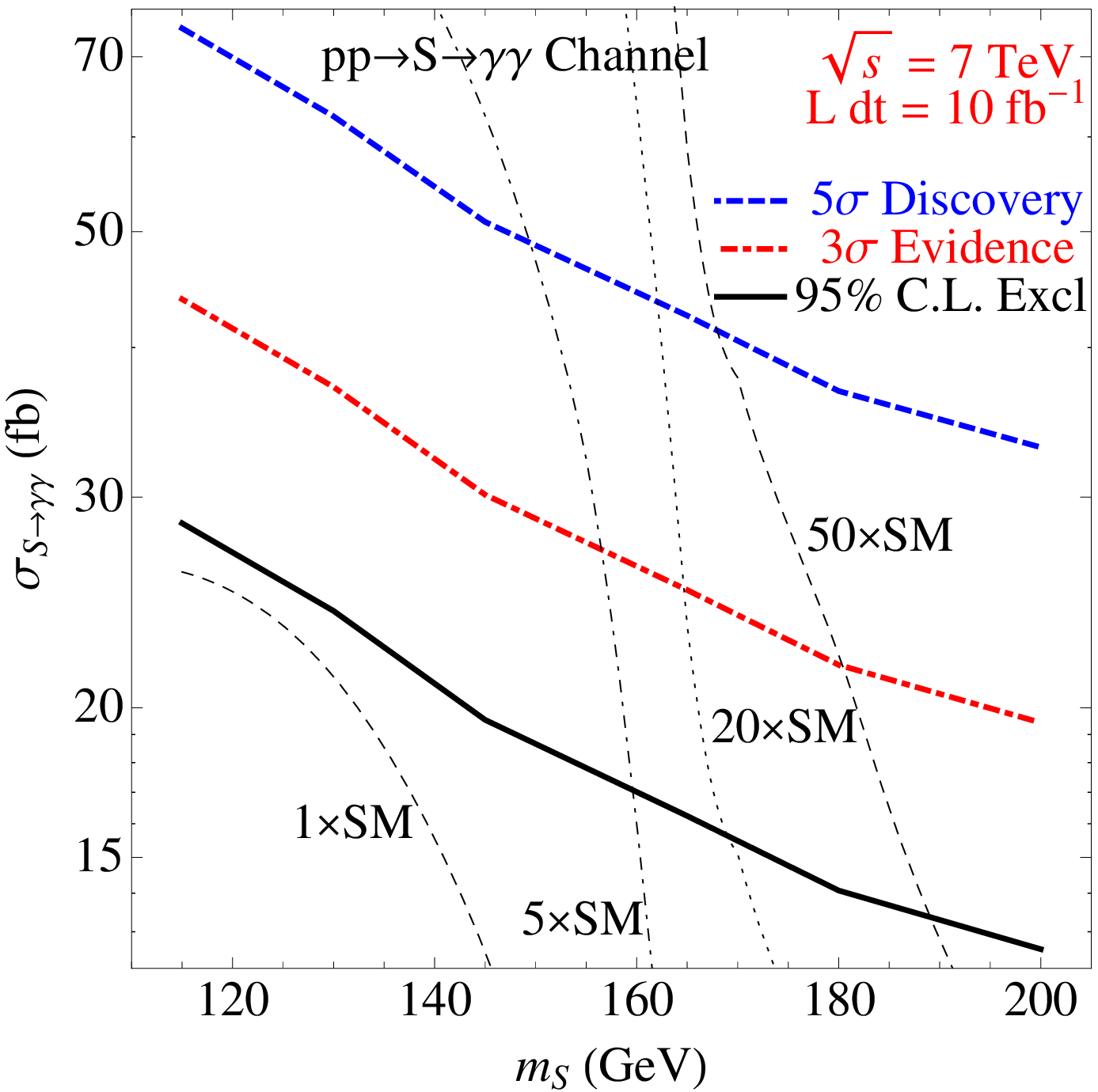}
\includegraphics[scale=0.5, angle=0]{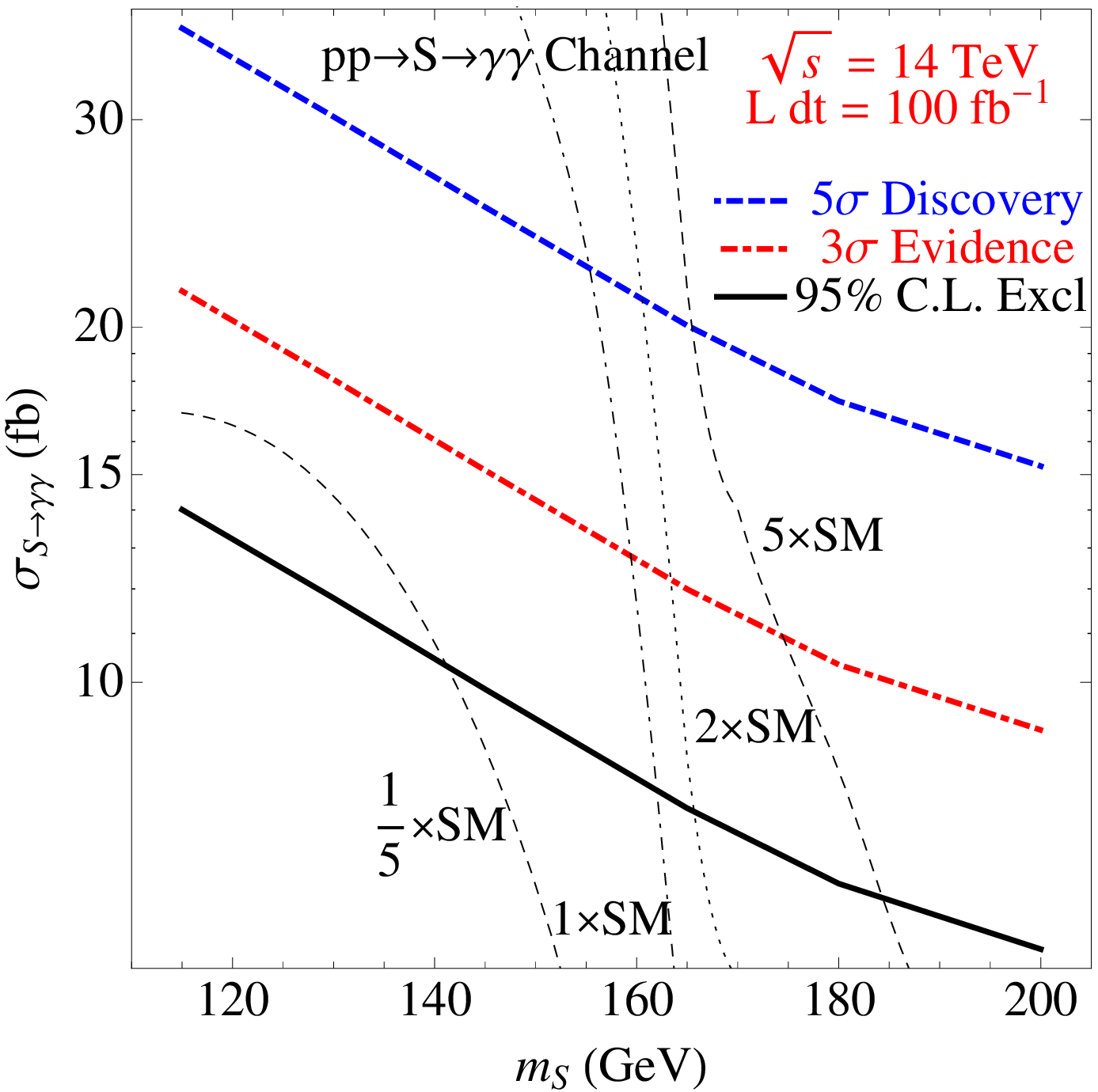}\\
\includegraphics[scale=0.5, angle=0]{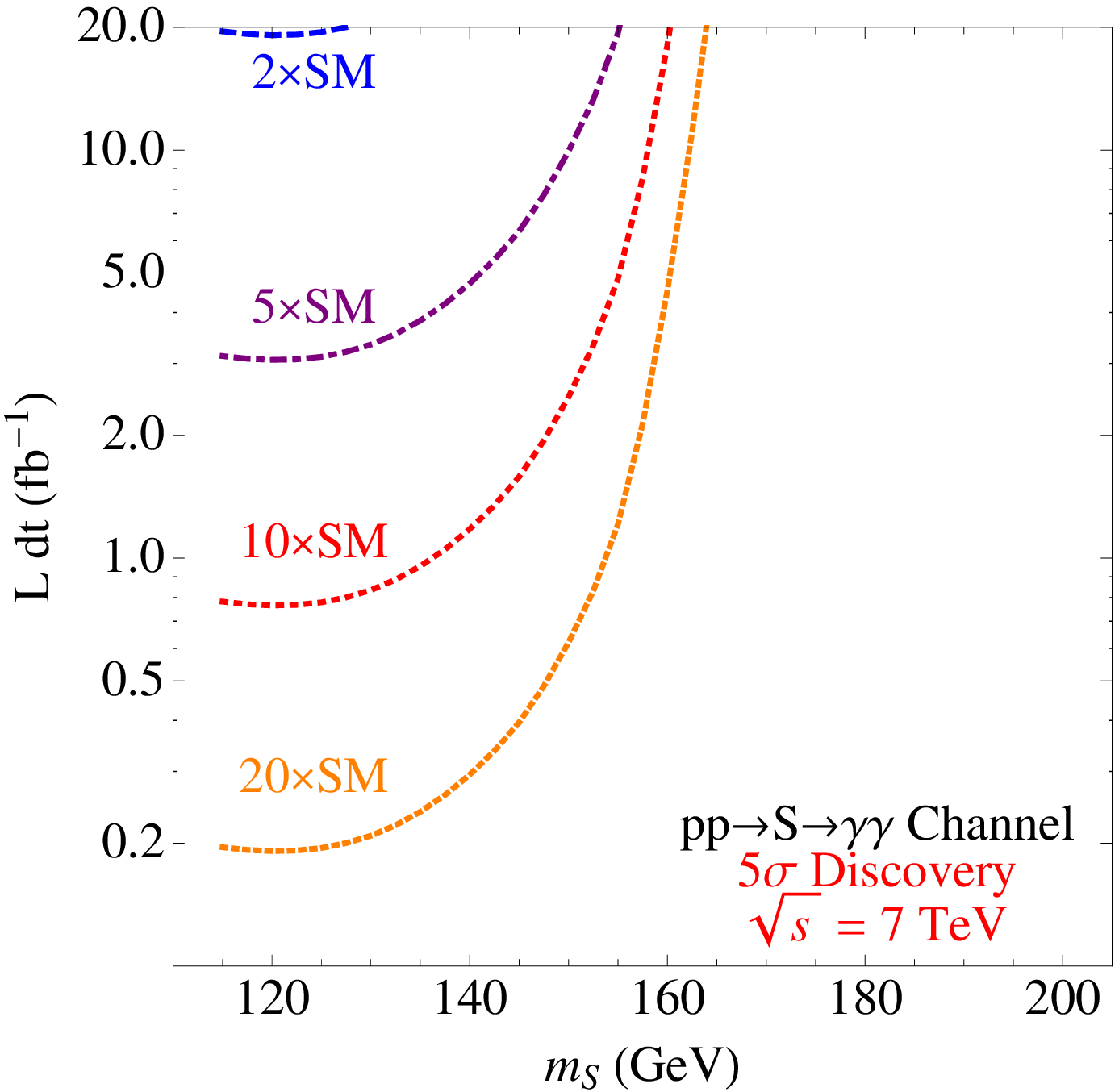}
\includegraphics[scale=0.5, angle=0]{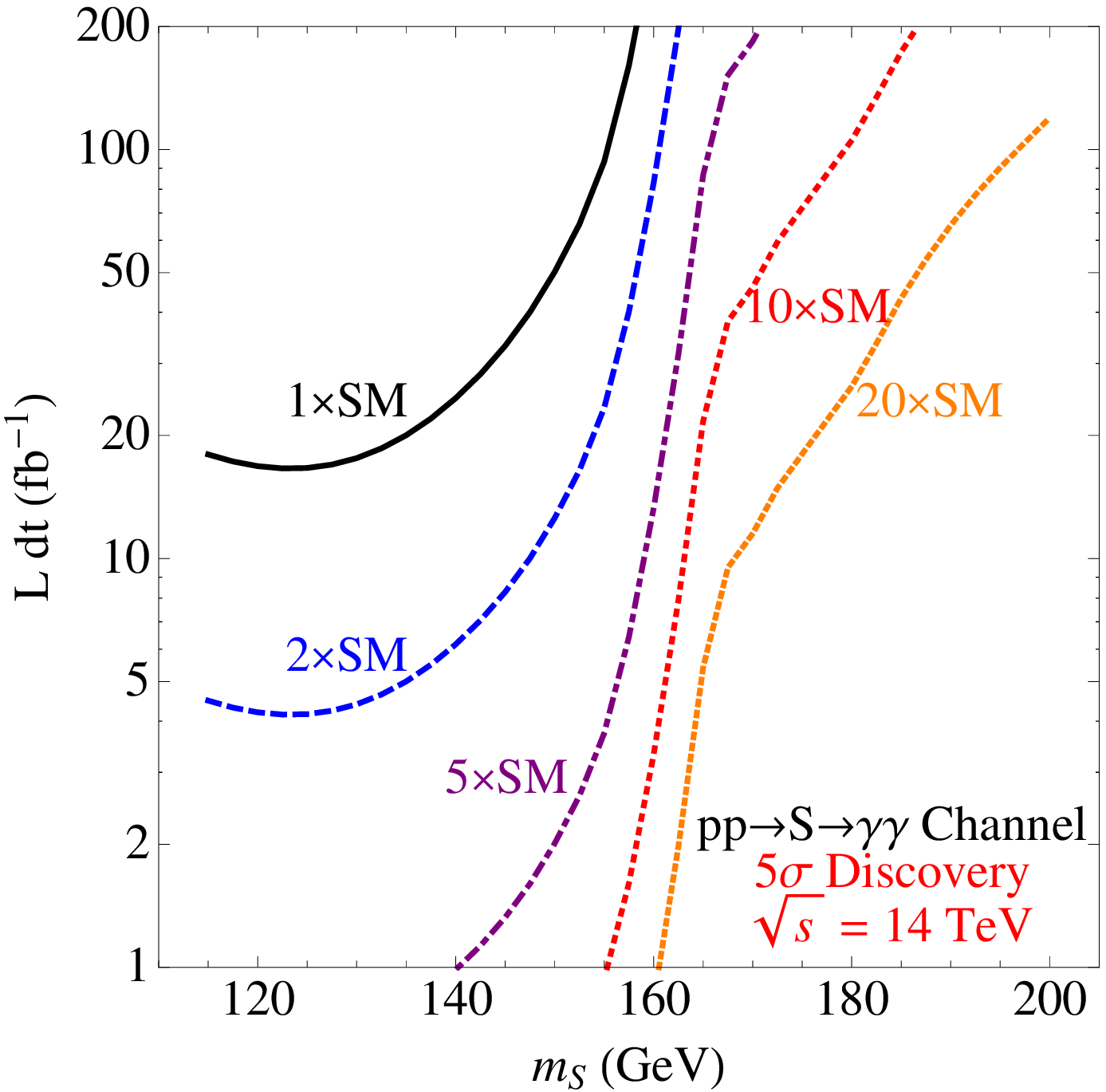}\\
\caption{LHC reach in the $\gamma \gamma$ channel for 7 TeV (left panels) and 14 TeV (right panels).  The top panels show the diphoton cross section reach for discovery, evidence and exclusion with multiples of the SM expectation for comparison.  The luminosity required for $5\sigma$ discovery with a multiple of the SM Higgs diphoton cross section is given in the lower panels.}
\label{fig:s2aareach}
\end{center}
\end{figure}

The overall reach of the LHC in the diphoton channel is summarized in Fig.~\ref{fig:s2aareach}.  The 7 TeV early LHC running is poised to discover scalars with ${\cal O}(5)$ times the SM rate of diphotons in the low mass range $m_S \lesssim 140$ GeV.  For smaller rates or higher masses the upgraded 14 TeV LHC has excellent discovery reach.

\subsection{$S\to Z\gamma$}

The $Z\gamma\to \ell^+\ell^-\gamma$ signature is more difficult to produce in the SM than $\gamma\gamma$ due to the weaker $Z$ boson coupling and small branching fraction to charged leptons.  However, the $\gamma\gamma$ backgrounds have a strong QCD component that is relatively absent in the $Z\gamma$ case, yielding an overall smaller background and hence greater sensitivity in the $Z\gamma$ channel.

\begin{table}[t]
\caption{Signal acceptance for trial scalar masses and cross section after level C3 and C4 cuts.}
\begin{center}
\begin{tabular}{|c|cc|}
\hline
Acceptance after C4 cut &$\sqrt s=7$ TeV&$\sqrt s=14$ TeV\\
\hline
$m_S=115$ GeV$/c^2$& 0.23& 0.20 \\
$m_S=130$ GeV$/c^2$& 0.31& 0.28\\
$m_S=165$ GeV$/c^2$& 0.33& 0.30\\
\hline\hline
Cross section after C3 cuts (fb) &$\sqrt s=7$ TeV&$\sqrt s=14$ TeV\\
\hline
$\ell^+\ell^- \gamma$& 673 & 1203\\
$Z+$jets& 17 & 45 \\
$t_\ell \bar t_\ell$&0.04 & 0.23 \\
\hline
Total Background (fb) & 690 & 1248\\
\hline
\end{tabular}
\end{center}
\label{tab:sigbkgza}
\end{table}%
For $pp\to S \to Z \gamma$, we generate events in MadEvent with an effective $hgg$ and $hZ\gamma$ coupling.  The irreducible background is the continuum $Z \gamma$ process while the reducible backgrounds are $Z j$, $Zjj$ and $t\bar t\to b\bar b \ell^+ \ell^- +\met$ with both sets generated in ALPGEN.  Following the ATLAS cuts~\cite{atlas-gamz} in the search for $Z\gamma$, we require (denoted `C3' cuts)
\begin{eqnarray}
 p_{T,\gamma} > 15\text{ GeV}, \quad p_T(\ell) &>& 20\text{ GeV},\\
  |\eta_\gamma,\ell| < 2.47, \quad  \quad \Delta R_{\gamma\ell} &>& 0.7, \quad \quad \Delta R_{\ell\ell} > 0.4,\\
|M_{\ell^+ \ell^-} - m_Z| &<& 5\text{ GeV}.
\end{eqnarray}
We require the $Z$ boson to be on-shell to reject $t\bar t$ backgrounds.  As in the $\gamma \gamma$ channel, we define the signal region to pass the following cut (denoted `C4' cuts) based on the trial scalar mass
\begin{equation}
|M_{\ell^+ \ell^-\gamma} - m_S|< 2.5\text{ GeV}.
\end{equation}
Given these cuts, we show, in Table~\ref{tab:sigbkgza}, the acceptance and cross section.  

\begin{figure}[t]
\begin{center}
\includegraphics[scale=0.5, angle=0]{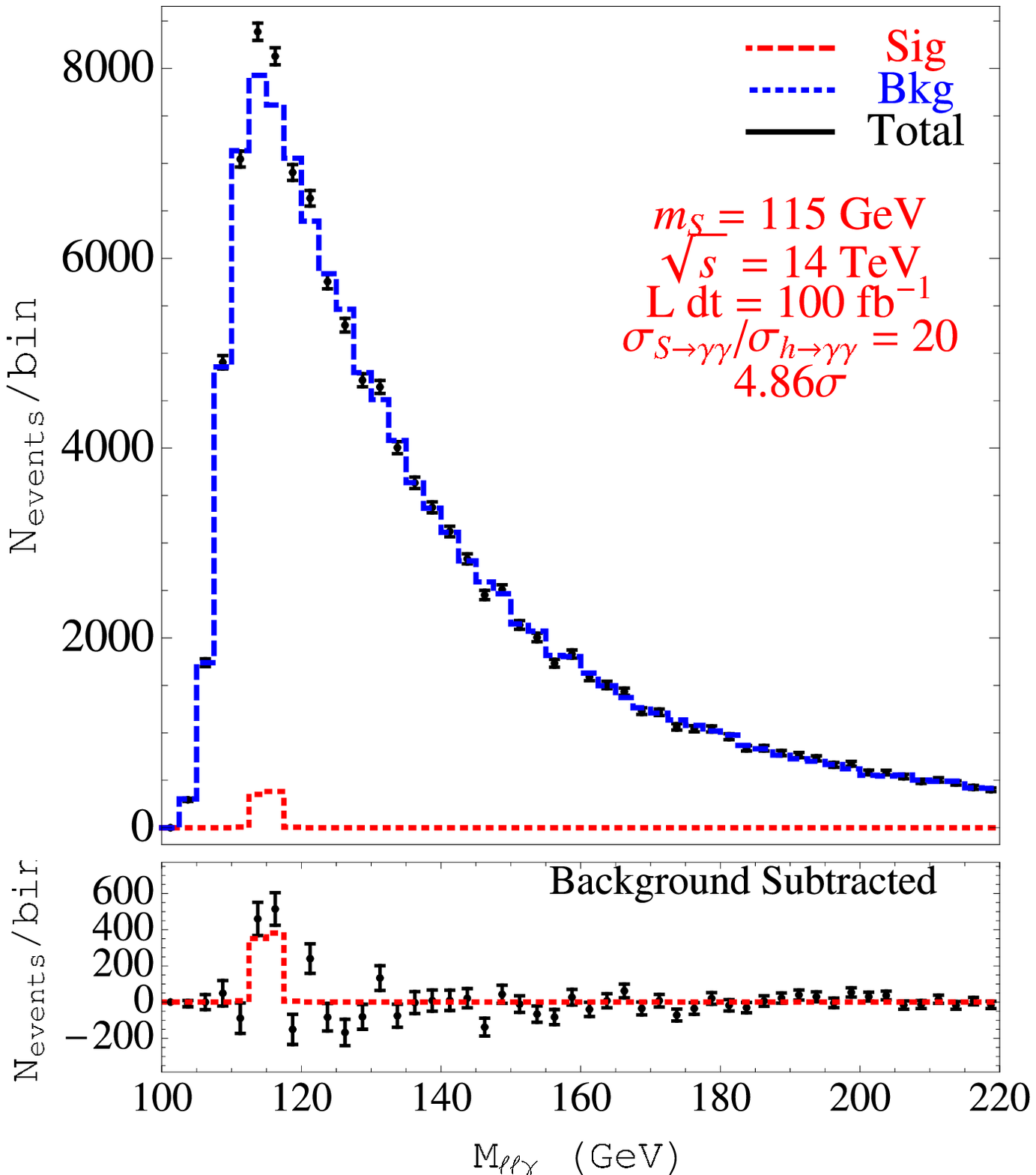}
\includegraphics[scale=0.5, angle=0]{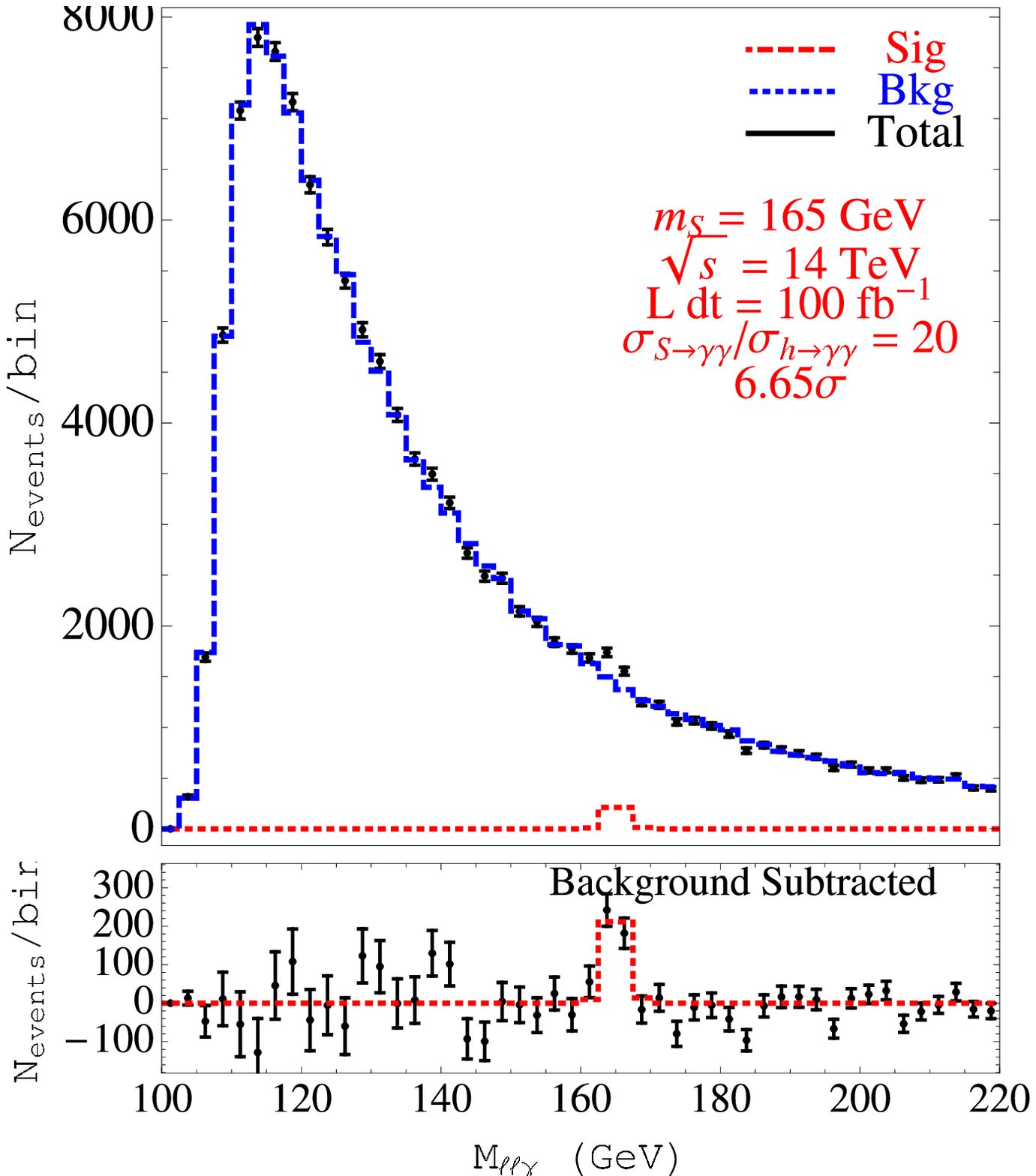}\\\vspace{.1in}
\caption{Distributions of $M_{\ell^+\ell^-\gamma}$ from pseudo-experiments for $m_S=115$ and 165 GeV$/c^2$ 
and $\sqrt s=14$ TeV for 100 fb$^{-1}$ of integrated luminosity, respectively.  
The background subtracted distribution more clearly shows the possible $\ell^+\ell^- \gamma$ peak.}
\label{fig:zgam}
\end{center}
\end{figure}

We present in Fig.~\ref{fig:zgam} the $M_{\ell^+\ell^-\gamma}$ distribution for the 14 TeV LHC with 100 fb$^{-1}$ of data.  Note that the $Z$ boson reconstruction paired with the minimum $p_T(\gamma)$ provides a cutoff in the $M_{\ell^+ \ell^-\gamma}$ distribution  near 105 GeV$/c^2$.  It's interesting that these cuts coupled with the natural decrease of the cross section at higher invariant mass scales creates a peak in the background in the $M_{\ell^+ \ell^-\gamma}\sim110-115$ GeV$/c^2$ range.  This may be problematic for searches of resonances in $Z\gamma$ in this mass range due to decreased acceptance, but may be alleviated with different threshold cuts.\footnote{This feature of $Z\gamma$ background also demonstrates the importance of understanding its leakage into the $\gamma\gamma$ background, especially in the 115 GeV/c$^2$ region.}  In addition, shape-based analyses will help further probe this region of scalar mass.  

\begin{figure}[t]
\begin{center}
\includegraphics[scale=0.5, angle=0]{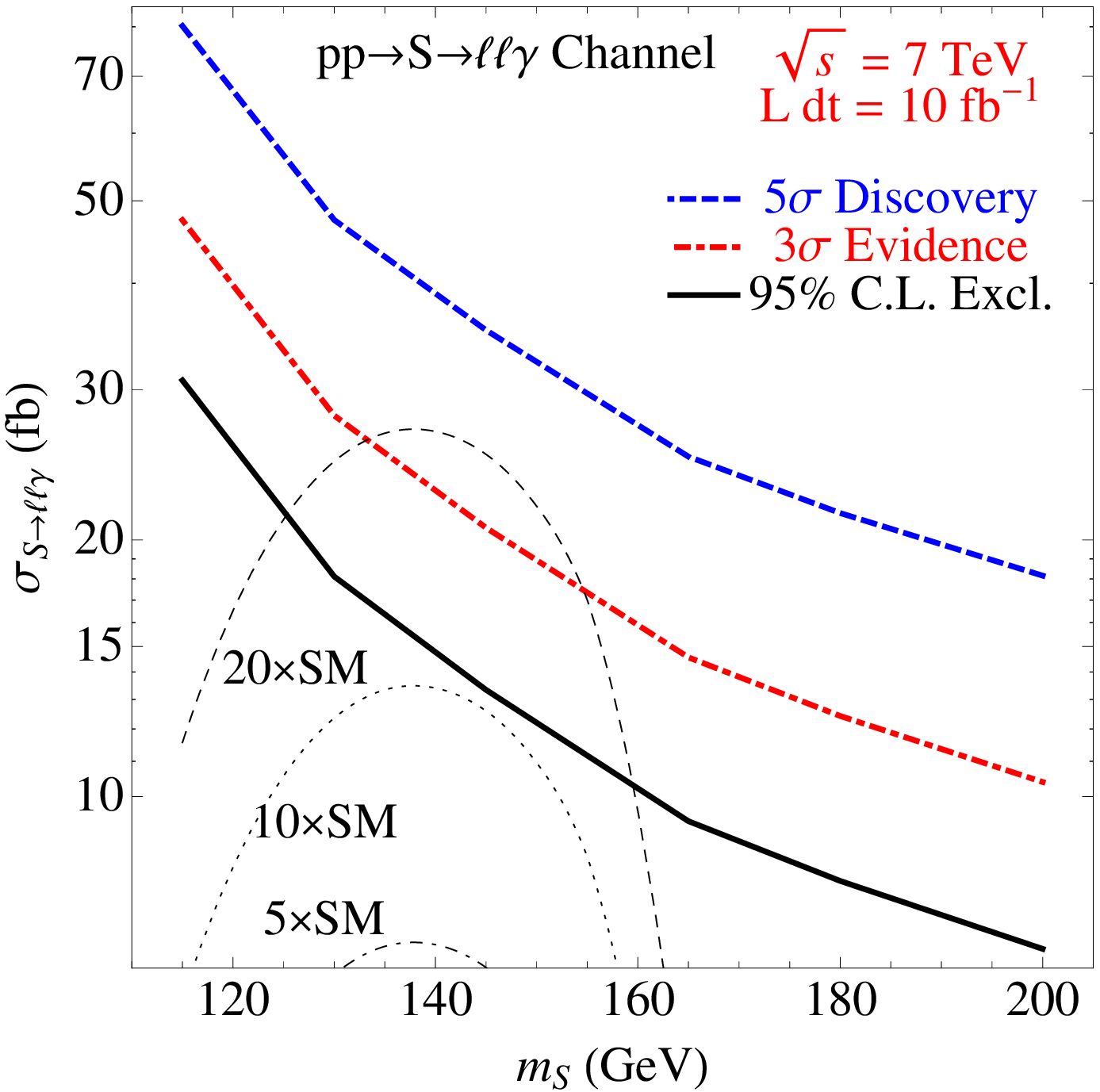}
\includegraphics[scale=0.5, angle=0]{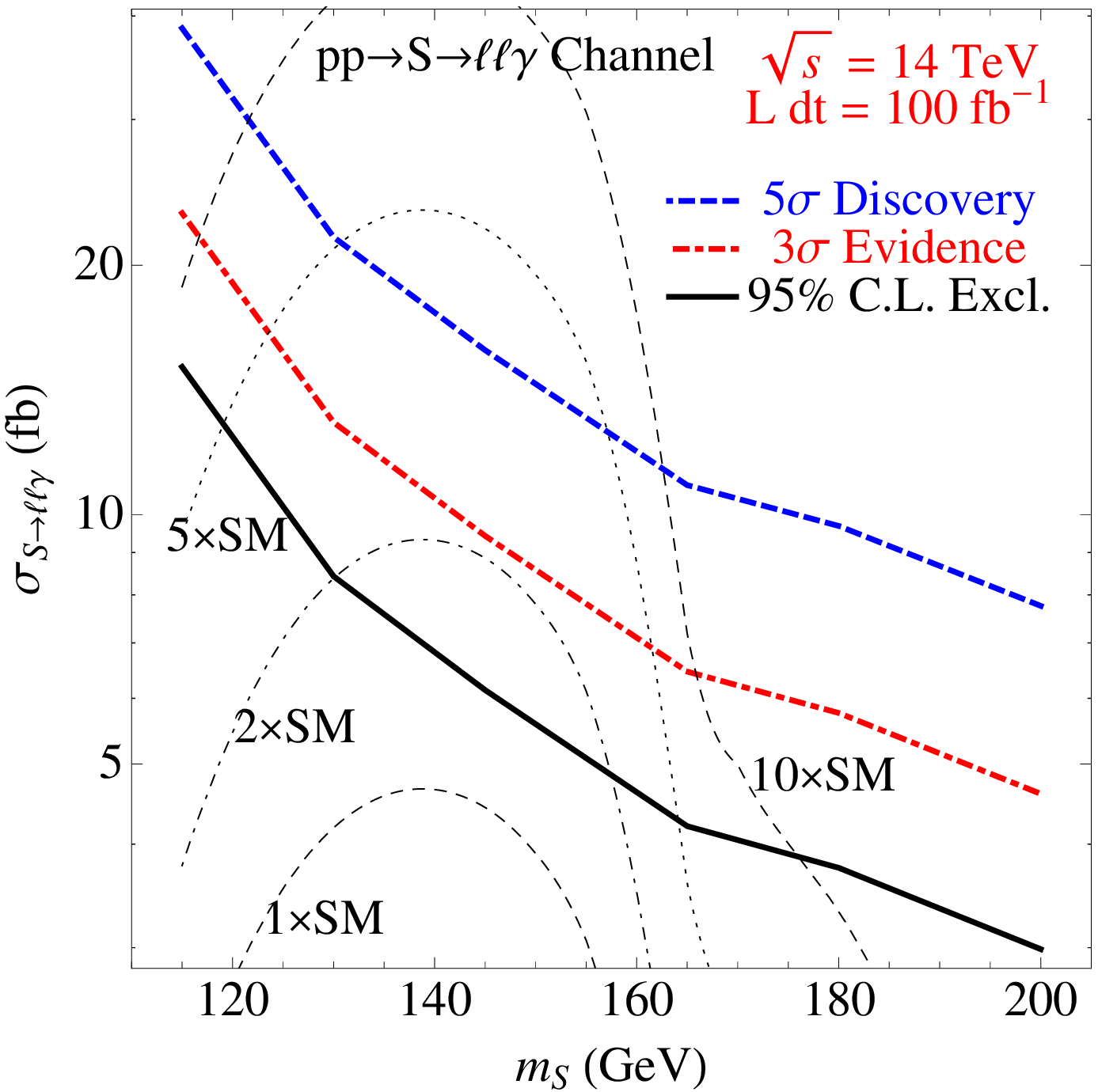}
\includegraphics[scale=0.5, angle=0]{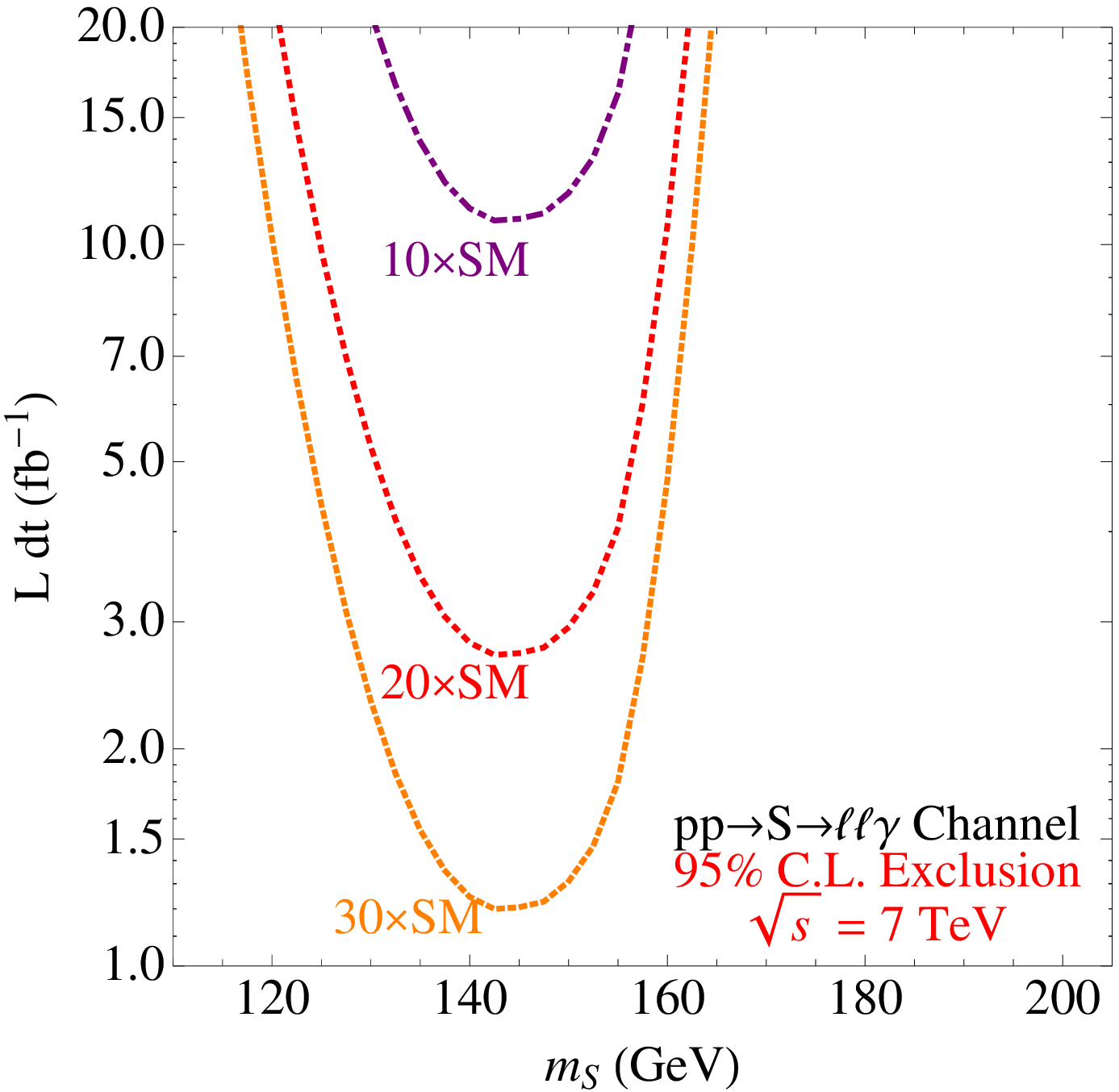}
\includegraphics[scale=0.5, angle=0]{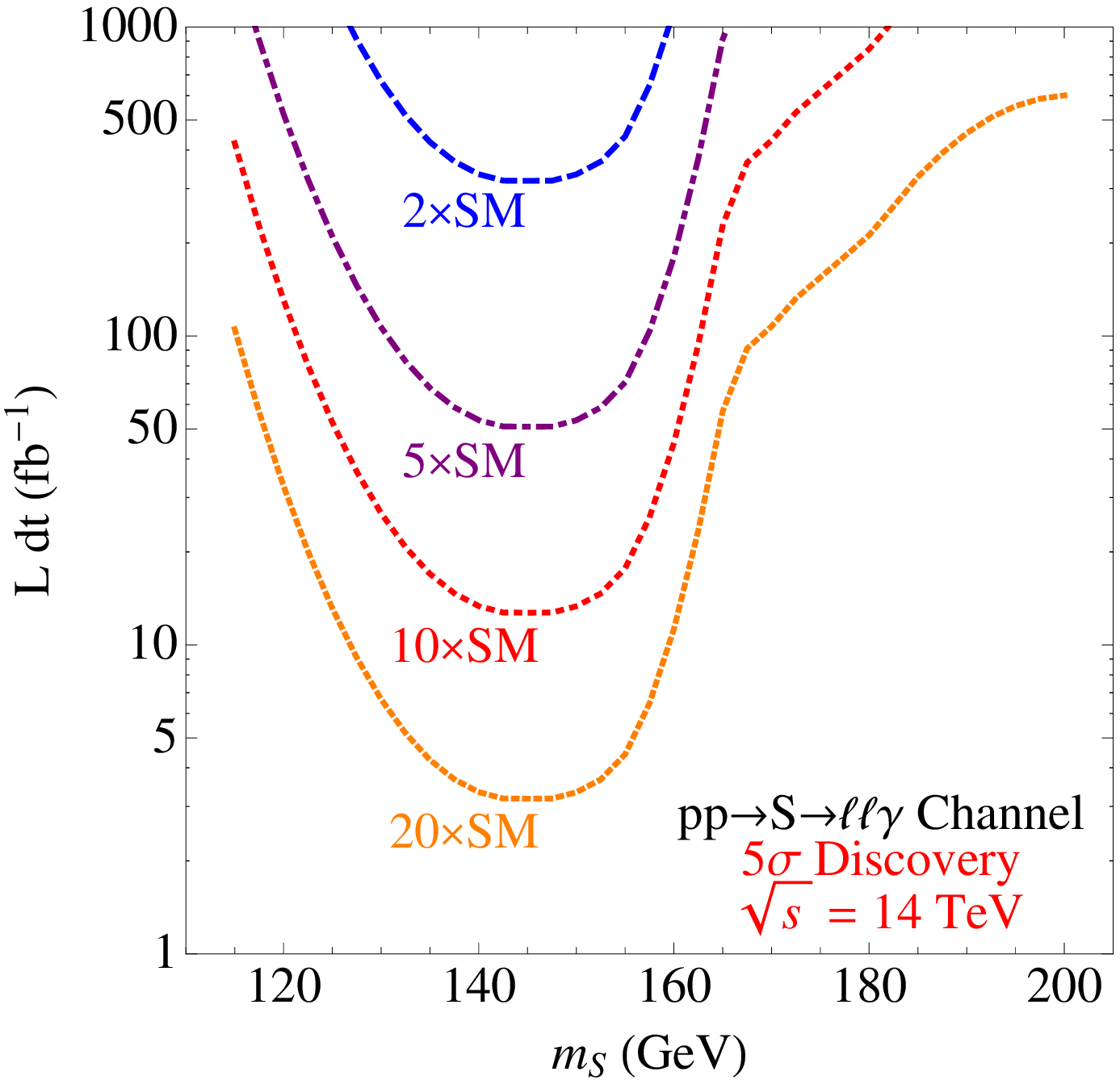}
\caption{LHC reach in the $\ell\ell\gamma$ channel for 7 TeV (left panels) and 14 TeV (right panels).  The top panels show the $\ell\ell\gamma$ cross section reach for discovery, evidence and exclusion with multiples of the SM expectation for comparison.  The luminosity required for various multiples of the SM Higgs $Z\gamma$ cross section is given in the lower panels.  We illustrate 95\% C.L. exclusion for the 7 TeV case and discovery for the 14 TeV case.}
\label{fig:s2zareach}
\end{center}
\end{figure}

In Fig.~\ref{fig:s2zareach}, we present the reach of the LHC in the $\ell\ell\gamma$ channel.  For the early running of the LHC at 7 TeV
the reach is limited to large enhancements of the $pp\to S\to \ell \ell \gamma$ cross section above the SM.  
This is largely due to the penalty paid by the $Z\to \ell\ell$ branching fraction.  At these energies, cross sections between 20 -- 70 fb can be discovered over the mass range we consider.  The maximum sensitivity in this channel for scalars with a fixed enhancement of production over the SM is $m_S\approx 140$ GeV.

\section{Conclusions}

In this work we considered the possibility of an electroweak singlet scalar as the Higgs imposter. Such a singlet scalar has loop-induced couplings to pairs of vector bosons arising only at the level of dimension five operators. We demonstrated that patterns of the singlet decay branching fractions into dibosons are generically very different from what would be expected of a SM Higgs. In particular, decay widths into $\gamma\gamma$ and $Z\gamma$ final states could naturally be enhanced, by orders of magnitude, over those of a SM Higgs. Therefore both channels could serve as the primary discovery channels for a Higgs imposter even for masses above $WW$ threshold.

\begin {acknowledgements}
The authors acknowledge inspiration from Stefani Germanotta, Tom LeCompte, Heidi Schellman, Michael Schmitt, and
Ciaran Williams, as well as Paddy Fox and Dave Tucker-Smith for explaining benefits of a Higgs friend.
This work was supported in part by the U.S. Department of Energy under
contracts No. DE-AC02-06CH11357 and No. DE-FG02-91ER40684. Fermilab is operated by the 
Fermi Reseacrh Alliance under contract DE-AC02-07CH11359 with
the U.S. Department of Energy.
\end{acknowledgements}

\end{document}